\documentclass[preprint,aps,12pt,showpacs,nofootinbib,tightenlines,amsmath,amssymb]{revtex4}
\usepackage{amsmath}
\usepackage{graphicx}
\usepackage{amssymb}
\usepackage{color}

\usepackage{CJK}
\usepackage{indentfirst}
\usepackage{bm}
\usepackage{mathtools}
\usepackage{amsmath}
\usepackage{graphicx}
\usepackage{epsfig}
\usepackage[retainorgcmds]{IEEEtrantools}
\DeclareMathOperator{\Tr}{Tr}
\usepackage{caption}
\newcommand{\sslash}{\!\!\!\slash}

\allowdisplaybreaks[1]

\begin{document}

\begin{center}
\LARGE\bf Gravity and Spin Forces \\ in Gravitational Quantum Field Theory
\end{center}

\begin{center}
\rm Yue-Liang Wu$^{\mathrm a,b,c)}$, \ \ Rui Zhang$^{\dag\mathrm a,b)}$
\end{center}

\begin{center}
\begin{footnotesize} \sl
${}^{\mathrm a)}$Key Laboratory of Theoretical Physics, Institute of Theoretical Physics,\\
Chinese Academy of Sciences, Beijing 100190, China \\
${}^{\mathrm b)}$ School of Physical Sciences, University of Chinese Academy of Sciences,\\
No. 19A Yuquan Road, Beijing 100049, China\\
${}^{\mathrm c)}$ International Center for Theoretical Physics Asia-Pacific (ICTP-AP), \\  University of Chinese Academy of Sciences,
Beijing 100049, China\\
\end{footnotesize}
\end{center}

\vspace*{2mm}

\begin{center}
\begin{minipage}{15.5cm}
\parindent 20pt\footnotesize
In the new framework of gravitational quantum field theory (GQFT) with spin and scaling gauge invariance developed  in Phys. Rev. D\textbf{93} (2016) 024012-1~\cite{Wu:2015wwa}, we make a perturbative  expansion for the full action in a background field which accounts for the early inflationary universe. We decompose the bicovariant vector fields of gravifield and spin gauge field with Lorentz and spin symmetries SO(1,3) and SP(1,3) in biframe spacetime into SO(3) representations for deriving the propagators of the basic quantum fields and extract their interaction terms. The leading order Feynman rules are presented. A tree-level 2 to 2 scattering amplitude of the Dirac fermions, through a gravifield and a spin gauge field, is calculated and compared to the Born approximation of the potential. It is shown that the Newton's gravitational law in the early universe is modified due to the background field. The spin dependence of the gravitational potential is  demonstrated.
\end{minipage}
\end{center}

\pacs{04.60.Bc, 04.20.Cv}
\maketitle

\section{Introduction}
\label{intro}

The gravitational quantum field theory (GQFT) with spin and scaling gauge invariance was developed in ~\cite{Wu:2015wwa,Wu:2015hoa} to overcome the long term obstacle between the general theory of relativity (GR) and quantum mechanics.  In fact, there has been enormous efforts on the theory beyond Einstein's theory since the GR was established by Einstein in 1915~\cite{Einstein:1915by}. The metric describing the geometry of the spacetime are commonly factorized linearly to explore the quantum structure of gravity and its interaction with matter fields~\cite{Voronov:1973kga,Choi:1994ax}, and the Ricci scalar has been shown to be the key of the dynamics of gravity. The property of GR with spin and torsion was investigated in Refs. ~\cite{Kerlick:1975tr,Hehl:1976kj,Hehl:1974cn}, where the totally antisymmetric coupling of the torsion to spin was presented. The general quadratic terms of the 2-rank tensor fields that satisfy the ghost-free and locality conditions were discussed in~\cite{VanNieuwenhuizen:1973fi}. With the tool named tensor projection operators developed in Ref.~\cite{Rivers:1964rj}, which projects the SO(1,3) tensor representation to the components of different SO(3) representations, the general propagators and gauge freedoms were investigated and extrapolated to a more general case including propagating torsion~\cite{Sezgin:1979zf}. The totally antisymmetric part and its renormalizability was anayzed in~\cite{Sezgin:1980tp}. 

Recently, a new framework of gravitational quantum field theory (GQFT) was proposed to treat the gravitational interaction on the same footing as electroweak and strong interactions~\cite{Wu:2015wwa,Wu:2015hoa}. Where a biframe spacetime is initiated, namely, the locally flat non-coordinate spacetime and the globally flat Minkowski spacetime, a basic gravifield is defined on the biframe spacetime as a bicovariant vector field which  is in general a 16-component field. The spin gauge field and scaling gauge field are introduced to keep the action invariant under a local $SP(1,3)\times SG(1)$ gauge transformation. A non-constant background solution has been obtained, which may account for the inflationary behaviour of the early universe. In a proceeding work, a more general action for a hyperunified field theory (HUFT) under the hyper-spin gauge and scaling gauge symmetries was proposed~\cite{Wu:2017urh} to merge all elementary particles into a single hyper-spinor field and unify all basic forces into a fundamental interaction governed by a hyper-spin gauge symmetry. A background solution remains to exist. In such a HUFT, it enables us to demonstrate the gravitational origin of gauge symmetry as the hyper-gravifield plays an essential role as a Goldstone-like field. The gauge-gravity and gravity-geometry correspondences lead to the gravitational gauge-geometry duality.  It has been shown that a general conformal scaling gauge symmetry in HUFT results in a general condition of coupling constants, which eliminates the higher derivative terms due to the quadratic Riemann and Ricci tensors, so that the HUFT will get rid of the so-called unitarity problem caused by the higher order gravitational interactions. To demonstrate explicitly, in the present paper, we consider the gravitational interactions of gravifield and spin gauge field only in four dimensional case with a background field solution. Expanding the full action under such a background field, it is natural to extract the dynamics and interactions of the quantum fields. The interactions among these fields will reflect the gravitational behavior in the early universe.

\section{Action expansion in a non-constant background field}

\label{expandaction}

Let us start from a basic action by simply taking four dimentional spacetime, i.e., D=4, from the hyperunified field theory (HUFT)~\cite{Wu:2017urh} in hyper-spacetime, 
\begin{align}
\label{action}
  I_H &=\int[dx]\chi L = \int[dx]\chi \left\{\frac{1}{2}(\hat{\chi}_{a}{}^{\mu}\bar{\Psi}\gamma^{a}{\mathrm i}D_{\mu}\Psi + H.C.)- \frac{1}{4}[\hat{\chi}^{\mu\nu\mu'\nu'}_{aba'b'}R^{ab}_{\mu\nu}R^{a'b'}_{\mu'\nu'}+ \hat{\chi}^{\mu\mu'}\hat{\chi}^{\nu\nu'}W_{\mu\nu}W_{\mu'\nu'}] \right.
 \nonumber \\
  &+\left.\frac{1}{4}\alpha_E \phi^2\hat{\chi}^{\mu\nu\mu'\nu'}_{aa'}G^{a}_{\mu\nu}G^{a'}_{\mu'\nu'}+ \frac{1}{2}\hat{\chi}^{\mu\nu}d_{\mu}\phi d_{\nu}\phi-\lambda_s\phi^4\right\}.
\end{align}
where $\chi_{\mu}{}^{a}$ is the gravifield, $\Omega_{\mu}{}^{ab}$ is the spin gauge field antisymmetric in $(a,b)$, $\hat{\chi}_{a}{}^{\mu}$ is the inverse of the gravifield, $\phi$ is the scalar field and $w_\mu$ is the scaling gauge field. And we have used the notations
\begin{align*}
  R_{\mu\nu}^{ab} =&\partial_\mu \Omega_{\nu}{}^{ab}-\partial_\nu \Omega_{\mu}{}^{ab}+\Omega_{\mu}{}^{a}{}_{c}\Omega_{\nu}{}^{cb}- \Omega_{\nu}{}^{a}{}_{c}\Omega_{\mu}{}^{cb},\qquad W_{\mu\nu}=\partial_{\mu}w_\nu-\partial_{\nu}w_\mu, \\
  G^{a}_{\mu\nu}=&( \partial_\mu \chi_\nu{}^{a}+\partial_\mu (ln\phi) \chi_\nu{}^{a}- \partial_\nu \chi_\mu{}^{a}-\partial_\nu (ln\phi) \chi_\mu{}^{a}),\\
  {\mathrm i}D_{\mu}=&{\mathrm i}\partial_{\mu}+\Omega{\mu}{}^{ab}\frac{1}{2}\Sigma_{ab},\qquad d_{\mu}=\partial_{\mu}-g_w w_\mu, \qquad \hat{\chi}^{\mu\nu}=\hat{\chi}_{a}{}^{\mu}\hat{\chi}_{b}{}^{\nu}\eta^{ab} .
  \end{align*}
The tensors are taken the general forms presented in~\cite{Wu:2017urh}    
  \begin{align*}
  \hat{\chi}^{\mu\nu\rho\sigma}_{abcd}=&g_1 \eta_{ac} \eta_{bd} \hat{\chi}^{\mu \rho} \hat{\chi}^{\nu \sigma} + \tfrac{1}{2} g_2 (\hat{\chi}_{d}{}^{\mu} \hat{\chi}_{c}{}^{\nu} \hat{\chi}_{b}{}^{\rho} \hat{\chi}_{a}{}^{\sigma} + \hat{\chi}_{c}{}^{\mu} \hat{\chi}_{d}{}^{\nu} \hat{\chi}_{a}{}^{\rho} \hat{\chi}_{b}{}^{\sigma})  
  \\&- \tfrac{g_1+g_2}{2}  \bigl(\eta_{bd} (\hat{\chi}_{c}{}^{\nu} \hat{\chi}_{a}{}^{\sigma} \hat{\chi}^{\mu \rho} + \hat{\chi}_{c}{}^{\mu} \hat{\chi}_{a}{}^{\rho} \hat{\chi}^{\nu \sigma}) + \eta_{ac} (\hat{\chi}_{d}{}^{\nu} \hat{\chi}_{b}{}^{\sigma} \hat{\chi}^{\mu \rho} + \hat{\chi}_{d}{}^{\mu} \hat{\chi}_{b}{}^{\rho} \hat{\chi}^{\nu \sigma})\bigr) 
  \\& -\tfrac{1}{2} g_4 (\hat{\chi}_{b}{}^{\sigma} \hat{\chi}_{a}{}^{\mu} \hat{\chi}_{d}{}^{\nu} \hat{\chi}_{c}{}^{\rho} + \hat{\chi}_{b}{}^{\mu} \hat{\chi}_{c}{}^{\nu} \hat{\chi}_{d}{}^{\rho} \hat{\chi}_{a}{}^{\sigma} + \hat{\chi}_{d}{}^{\mu} \hat{\chi}_{a}{}^{\nu} \hat{\chi}_{b}{}^{\rho} \hat{\chi}_{c}{}^{\sigma} + \hat{\chi}_{c}{}^{\mu} \hat{\chi}_{b}{}^{\nu} \hat{\chi}_{a}{}^{\rho} \hat{\chi}_{d}{}^{\sigma})
  \\&+  \tfrac{1}{2} g_4 \bigl(\eta_{bd} (\hat{\chi}_{a}{}^{\nu} \hat{\chi}_{c}{}^{\sigma} \hat{\chi}^{\mu \rho} + \hat{\chi}_{a}{}^{\mu} \hat{\chi}_{c}{}^{\rho} \hat{\chi}^{\nu \sigma}) + \eta_{ac} (\hat{\chi}_{b}{}^{\nu} \hat{\chi}_{d}{}^{\sigma} \hat{\chi}^{\mu \rho} + \hat{\chi}_{b}{}^{\mu} \hat{\chi}_{d}{}^{\rho} \hat{\chi}^{\nu \sigma})\bigr)
  \\
  \hat{\chi}^{\mu\nu\rho\sigma}_{ab}=& \eta_{ab} \hat{\chi}^{\mu \rho} \hat{\chi}^{\nu \sigma} + 2  \hat{\chi}_{a}{}^{\sigma} \hat{\chi}_{b}{}^{\nu} \hat{\chi}^{\mu \rho} - 4  \hat{\chi}_{a}{}^{\nu} \hat{\chi}_{b}{}^{\sigma} \hat{\chi}^{\mu \rho}  .
\end{align*}
In the unitary basis $\chi \equiv \det \chi_{\mu}^{\; a} =1$, the background field solution in the unitary basis is found to be ~\cite{Wu:2015wwa}
\begin{align*}
  \bar{\chi}_\mu{}^a=\eta_\mu^a,\qquad \bar{\phi}^2 =\frac{6 \alpha_E}{\lambda_s}\frac{ \kappa_\mu\kappa^\mu }{ (1-\kappa\cdot x)^2},\qquad
  \bar{\Omega}_\mu{}^{ab}=0,\qquad \bar{w}_\mu=\frac{1}{g_w}\partial_\mu ln\bar{\phi}\, .
\end{align*}
The quantized field are expressed as:
\begin{align*}
  &\chi_\mu{}^a=\eta_\mu^a+h_\mu^a/M_W,\qquad \hat{\chi}_a{}^\mu=\eta_a^\mu-h_a^\mu/M_W+O(1/M_W^2),
  \\& \phi \to \bar{a}M_S+\phi,\qquad  \Omega_\mu{}^{ab}\to \bar{\Omega}_\mu{}^{ab}+\Omega_\mu{}^{ab},\qquad w_\mu\to \bar{w}_\mu+ w_\mu.
\end{align*}
with $\bar{\phi} = \bar{a} M_s$ and $\alpha_E M_s^2=\tfrac{1}{2}M_W^2$.
\paragraph{}
We can expand the action \eqref{action} and collect the leading order interactions and quadratic terms. As the quadratic term of the quantum gravifield includes a non-constant coefficient $\bar{a}(x)$, it is useful to absorb it into the field via a field-redefinition 
\begin{equation*}
h_\mu{}^a\rightarrow h_\mu{}^a/\bar{a}(x).
\end{equation*}
The final quadratic terms are given by:
\begin{align}\label{quadratic}
  &- \tfrac{12\alpha_E-1}{2} \partial_{a}\phi \partial^{a}\phi + 2\sqrt{2\alpha_E} \partial_{a}\phi \partial_{b}h^{ab} -  \tfrac{1}{2} \partial_{b}h^{c}{}_{c} \partial^{b}h^{a}{}_{a} - 2 \partial_{b}\phi \partial^{b}h^{a}{}_{a} -  \tfrac{1}{2} \partial_{b}h^{ab} \partial_{c}h_{a}{}^{c} + \partial^{b}h^{a}{}_{a} \partial_{c}h_{b}{}^{c} \nonumber
  \\&-   \partial_{a}h_{bc} \partial^{c}h^{ab} -  \tfrac{1}{2} \partial_{a}h_{cb} \partial^{c}h^{ab} + \frac{1}{2} \partial_{a}h_{cb} \partial^{c}h^{ab} + \frac{1}{2} \partial_{b}h_{ac} \partial^{c}h^{ab} + \tfrac{1}{2} \partial_{c}h_{ab} \partial^{c}h^{ab} -  \frac{1}{2} \partial_{c}h_{ab} \partial^{c}h^{ab} \nonumber
  \\&+ \frac{1}{2} \partial_{c}h_{ba} \partial^{c}h^{ab}  - \tfrac{1}{2} g_1 \partial_{d}\Omega_{abc} \partial^{d}\Omega^{abc} +  \tfrac{1}{2} g_1 \partial_{a}\Omega_{dbc} \partial^{d}\Omega^{abc} +  \tfrac{1}{2} g_1 \partial_{d}\Omega_{bac} \partial^{d}\Omega^{abc} +  g_1 \partial_{a}\Omega_{bcd} \partial^{d}\Omega^{abc}  \nonumber \\
&  +  \tfrac{1}{2} (g_1+g_2) \partial_{c}\Omega_{abd} \partial^{d}\Omega^{abc} -  g_2 \partial_{c}\Omega_{bad} \partial^{d}\Omega^{abc} 
 +  g_2 \partial_{a}\Omega_{bcd} \partial^{d}\Omega^{abc} +  \tfrac{1}{2} g_2 \partial_{d}\Omega_{bac} \partial^{d}\Omega^{abc} \nonumber \\
& - \tfrac{1}{2} g_4 \partial_{b}\Omega^{d}{}_{cd} \partial^{c}\Omega^{a}{}_{a}{}^{b} +  \tfrac{1}{2} g_4 \partial_{c}\Omega^{d}{}_{bd} \partial^{c}\Omega^{a}{}_{a}{}^{b} +  \tfrac{1}{2} g_4 \partial_{b}\Omega^{abc} \partial_{d}\Omega_{ac}{}^{d} + g_4
\partial^{c}\Omega^{a}{}_{a}{}^{b} \partial_{d}\Omega_{bc}{}^{d} \nonumber \\
& - \tfrac{1}{2} g_4 \partial_{b}\Omega^{abc} \partial_{d}\Omega_{ca}{}^{d} 
- g_4 \partial^{c}\Omega^{a}{}_{a}{}^{b} \partial_{d}\Omega_{cb}{}^{d}  + \tfrac{1}{2} \partial_{a}w_{b} \partial^{b}w^{a} -  \tfrac{1}{2} \partial_{b}w_{a} \partial^{b}w^{a}
\end{align}
There are other terms which involves two quantum fields, but with higher orders of the background field, we present them in the Appendix A. In the early universe, the background field $\bar{\phi}(x)$ is sufficiently small, so that we can ignore the effect of those terms and only consider the quadratic terms in~\eqref{quadratic}. 
\paragraph{}
Though the propagators can hardly be read from the action, we can utilize the tensor projection operators to decompose the spin components of the tensor fields, and then derive their propagators. The scaling gauge field decouples from the Dirac spinors, so we would not include it in our present considerations. We shall discuss the details in Sec.~\ref{projection}~~ We can also get the leading-order interaction terms which are given in the appendix B. Notice that we have absorbed the gauge coupling constant $g_h$, which depends on the normalization of coefficients $g_1$, $g_2$ and $g_4$. We shall do a field redefinition after some normalization of the propagator in Sec.~\ref{sec:spin_gauge_int} ~~ and turn the interactions to a usual form of gauge interactions.

\section{Tensor projection operators and propagators of gravifield and spin gauge field as well as scalar field}
\label{projection}
The SO(1,3) tensor-like fields $h_{\mu a}$ and $\Omega_{\mu ab}$ can be decomposed into different SO(3) spin-parity components:
\begin{align}\label{decompose}
  \phi&=0^+,\qquad h_{\mu a}=2^+\oplus 1^+\oplus 1^-\oplus 1^-\oplus 0^+\oplus 0^+\nonumber\\
  \Omega^{\mu ab}&=2^-\oplus 2^+\oplus 1^+\oplus 1^-\oplus 1^-\oplus 1^+\oplus 0^-\oplus 0^+  .
\end{align}
Following ref. ~\cite{VanNieuwenhuizen:1973fi}, we shall define the tensor projection operators $P^{J^P}_{f_1 f_2 }$, where the subscripts $f_1$ and $f_2$ denoting the field type, the superscripts $J$ and $P$ label the spin and parity. The tensor projection operators satisfy the following relations:
\begin{equation}\label{relation}
P^{J^P}_{f_1 f_2}P^{{J'}^{P'}}_{f'_1 f'_2}=P^{J^P}_{f_1 f'_2}\delta_{JJ'}\delta_{PP'}\delta_{f_2 f'_1} .
\end{equation}
with the definition
\begin{equation}
 P^{J^P}_{f_1 f_1}\equiv P^{J^P}_{f_1} .
\end{equation}
To be specific, we write down the explicit forms for the tensor projection operator of the $2^+$ component of the gravifield $h_\mu{}^a$
\begin{equation}
(P_{h}^{2^+})_{\mu a,\nu b}= \tfrac{1}{2} \theta^{\mu \nu} \theta^{ab}  + \tfrac{1}{2} \theta^{a\nu} \theta^{b\mu}  -  \tfrac{1}{3} \theta^{a\mu} \theta^{b\nu}
\end{equation}
with the definition 
\begin{equation*}
\theta^{\mu\nu}\equiv\eta^{\mu\nu}-\frac{p^\mu p^\nu}{p^2},\quad \omega^{\mu\nu}\equiv\frac{p^\mu p^\nu}{p^2}, 
\end{equation*}
and the tensor projection operator of the totally antisymmetric part of the spin gauge field $\Omega_\mu{}^{ab}$,
\begin{align*}
(P_{\Omega_1}^{1^+})_{\mu ab,\nu cd}&= \frac{1}{6} [ \theta^{\mu \nu}  (\theta^{bd} \omega^{ac}  -  \theta^{bc}  \omega^{ad} )  - \theta^{b\nu} \theta^{\mu d} \omega^{ac} + \theta^{b\nu} \theta^{\mu c} \omega^{ad} -  \theta^{bd} \theta^{\mu c} \omega^{a\nu} + \theta^{bc} \theta^{\mu d} \omega^{a\nu} \nonumber\\
&  +\theta^{\mu \nu} (\theta^{ac}  \omega^{bd} -  \theta^{ad} \omega^{bc}) + \theta^{a\nu} \theta^{\mu d} \omega^{bc}  -  \theta^{a\nu} \theta^{\mu c} \omega^{bd} + + \theta^{ad} \theta^{\mu c} \omega^{b\nu} -  \theta^{ac} \theta^{\mu d} \omega^{b\nu} \nonumber\\
&-  \theta^{a\nu} \theta^{bd} \omega^{\mu c} + \theta^{ad} \theta^{b\nu} \omega^{\mu c} + \theta^{a\nu} \theta^{bc} \omega^{\mu d} -  \theta^{ac} \theta^{b\nu} \omega^{\mu d} -  \theta^{ad} \theta^{bc} \omega^{\mu \nu} + \theta^{ac} \theta^{bd} \omega^{\mu \nu}]\\
(P_\Omega^{0^-})_{\mu ab,\nu cd}&= \frac{1}{6} [ \theta^{\mu\nu} (\theta^{ac} \theta^{bd} - \theta^{ad} \theta^{bc} ) + \theta^{\mu c} \theta^{ad} \theta^{b\nu} - \theta^{\mu c} \theta^{a\nu} \theta^{bd} - \theta^{\mu d} \theta^{ac} \theta^{b\nu}+ \theta^{\mu d} \theta^{a\nu} \theta^{bc} ] .
\end{align*}
The explicit forms of other tensor projection operators are presented in the appendix C.

In general, the tensor projection operators have the following properties, 
\begin{align}\label{operators}
 &P^{0^+}_\phi=1;\qquad P^{0^+}_{h_1}+P^{0^+}_{h_2}+P^{1^+}_h+P^{1^-}_{h_1}+P^{1^-}_{h_2}+P^{2^+}_h=\eta_{\mu\mu'}\eta_{aa'};\nonumber\\
 &P^{0^+}_\Omega+P^{0^-}_\Omega+P^{1^+}_{\Omega_1}+P^{1^+}_{\Omega_2}+P^{1^-}_{\Omega_1}+P^{1^-}_{\Omega_2}+P^{2^-}_{\Omega}+P^{2^+}_\Omega=\frac{1}{2}\eta_{\mu\mu'}(\eta_{aa'}\eta_{bb'}-\eta_{ba'}\eta_{ab'}). 
\end{align}
\paragraph{}
Thus we can write the quadratic terms of the action in terms of the tensor projection operators as follows,
\begin{align}\label{decomposedqdterms}
&p^2\left\{ \phi(\frac{1-12\alpha_E}{2}P^{0^+}_\phi)\phi+\phi(-2\sqrt{6\alpha_E}P^{0^+}_{\phi h_1})_{\mu a}h^{\mu a} + h^{\mu a}\left(-P^{0^+}_{h_1} - 0P^{0^+}_{h_2} - \frac{\sqrt{3}}{2}P^{0^+}_{h_2 h_1} + \frac{\sqrt{3}}{2}P^{0^+}_{h_1 h_2}\right. \right.\nonumber
\\&\left.  0P_{h_1}^{1^-} + 0 P_{h_2}^{1^-} + \frac{1}{2}P^{1^-}_{h_2 h_1} + (-\frac{1}{2})P^{1^-}_{h_1 h_2} + 0 P^{1^+}_{h}  + \frac{1}{2}P^{2^+}_{h}\right)_{\mu a,\nu b}h^{\nu b} + \Omega^{\mu ab}\left( - \frac{g_1-g_2}{4}P^{2^-}_{\Omega} \right. \nonumber \\
& \left. -\frac{g_1 - g_2 + g_4}{3}P^{1^+}_{\Omega_1} +\frac{g_2+5g_1+2g_4}{12}\sqrt{2}P^{1^+}_{\Omega_1\Omega_2} - \frac{5g_2+g_1-2g_4}{12}\sqrt{2}P^{1^+}_{\Omega_2\Omega_1}+ \frac{g_2-g_1-g_4}{6}P^{1^+}_{\Omega_2} \right. \nonumber 
\\ & \left. +\frac{g_2-g_1-2g_4}{4}P^{1^-}_{\Omega_1} - \frac{\sqrt{2}g_4}{4}P^{1^-}_{\Omega_2\Omega_1}+\frac{\sqrt{2}g_4}{4}P^{1^-}_{\Omega_1\Omega_2}-\frac{g_2+2g_1}{2}P^{0^-}_{\Omega} \right. \nonumber \\
& \left. \left. + 0 P_{\Omega}^{0^+} + 0 P_{\Omega_2}^{1^-} + 0 P_{\Omega}^{2^+}  \right)_{\mu ab,\nu cd}\Omega^{\nu cd}  \right\}
\end{align}

The field equations of the field type ${\cal F}_{f_1}^{\, \{\mu\} }$ can be expressed by tensor projection operators as:
\begin{equation}\label{field_eq}
\sum_{f_2} a^{J^P}_{f_1 f_2 } (P^{J^P}_{f_1 f_2})_{\{\mu\},\{\nu\}} {\cal F}_{f_2}{}^{\{\nu\}} =( P^{J^P}_{f_1})_{\{\mu\},\{\mu'\}}\mathcal{J}_{f_1}^{\{\mu'\}}
\end{equation}
where $\mathcal{J}_{f_1}^{\{\mu'\}}$ is the corresponding source of the field ${\cal F}_{f_1}^{\{\mu\}}$. $a^{J^P}_{f_1 f_2}$  is the coefficient matrix of the field equations which are derived from~\eqref{decomposedqdterms}. We have used the relation in Eq.\eqref{relation} to obtain the above field equations.  Thus the propagators can be obtained by multiplying the operators $\sum_{f_4,f_1}(a^{J^P})^{-1}_{f_3 f_4 } (P^{J^P}_{f_3 f_4})$ on the left-hand side of ~\eqref{field_eq}
\begin{align}
&{\cal F}_{f_1}{}_{\{\mu\} } =\sum_{J,P,f_2} (a^{J^P})^{-1}_{f_1 f_2 } (P^{J^P}_{f_1 f_2})_{\{\mu\},\{\nu\}}\mathcal{J}_{f_2}^{\{\nu\}},
\nonumber\\
&(\Delta_{f_1f_2}^{-1})_{\{\mu\},\{\nu\}}=\frac{\delta {\cal F}_{f_1}{}_{\{\mu\} } }{\delta\mathcal{J}_{f_2}^{\{\nu\}}}=\sum_{J,P} (a^{J^P})^{-1}_{f_1 f_2 } (P^{J^P}_{f_1 f_2})_{\{\mu\},\{\nu\}}.
\end{align}
The explicit forms of the coefficient matrices are given by,
\begin{align}
a^{0^-}_{f_1 f_2} & =-(2g_1+g_2)p^2,\quad f_1=f_2=\Omega \label{0minus}
 \\
\nonumber \\
a^{0^+}_{f_1 f_2} & =\left(
\begin{array}{cccc}
0& 0& 0&  0\\
0& -2p^2 & 0& -2\sqrt{6\alpha_E}p^2\\
0& 0& 0& 0\\
0& -2\sqrt{6\alpha_E}p^2& 0& 1-12\alpha_Ep^2\\
\end{array} \right), \quad f_1,f_2 =(\Omega,h_1,h_2,\phi)\label{0plus}
\\
\nonumber \\
a^{1^-}_{f_1 f_2} &=\left(
\begin{array}{cccc}
\tfrac{g_2-g_1-2g_4}{2}p^2& 0& 0& 0\\
0& 0& 0& 0\\
0& 0& 0& 0\\
0& 0& 0& 0\\
\end{array} \right), \quad f_1,f_2 =(\Omega_1,\Omega_2,h_1,h_2) \label{1minus}
\\
\nonumber \\
a^{1^+}_{f_1 f_2} &=\left(
\begin{array}{ccc}
\tfrac{2g_2-2g_1-2g_4}{3}p^2& -\tfrac{\sqrt{2}(g_2-g_1-g_4)}{3}p^2 &0\\
-\tfrac{\sqrt{2}(g_2-g_1-g_4)}{3}p^2& \tfrac{g_2-g_1-g_4}{3}p^2 &0\\
0 & 0 & 0\\
\end{array}\right),  \quad f_1,f_2  =(\Omega_1,\Omega_2,h)\label{1plus}
\\
\nonumber \\
a^{2^-}_{f_1 f_2} & =  \frac{g_2-g_1}{4}p^2, \quad f_1=f_2 =\Omega \label{2minus}
\\
\nonumber \\
a^{2^+}_{f_1 f_2} & = \left(
\begin{array}{cc}
0& 0\\
0& p^2\\
\end{array}\right), \quad f_1,f_2 =(\Omega,h) \label{2plus}
\end{align}
It is obvious that most of the matrices are degenerate, and these degeneracies indicate certain symmetries of the quadratic terms~\cite{Sezgin:1979zf} relevant to unphysical degrees of freedom. When considering only the tree-level calculations, we do not need to know the exact gauge-fixing terms and gauge transformations by introducing the Faddeev-Popov ghosts. Instead, we can just apply the specific gauge-fixing conditions by setting the constraints 
\begin{align}
P_{h_2}^{0^+}h=P_{h}^{1^+}h=P_{h_1}^{1^-}h=P_{h_2}^{1^-}h=P_{\Omega}^{0^+}\Omega=P_{\Omega_2}^{1^-}\Omega=P_{\Omega_2}^{1^+}\Omega=P_{\Omega}^{2^+}\Omega=0
\end{align}
 without breaking the field equations, and neglect the corresponding lines in the coefficient matrices. Thus we only need to invert the ``reduced" matrices and get the propagators.
\paragraph{}
 The resulting propagators are given as follows in the specific gauge:
\begin{align}\label{prop}
  h^{\mu a}-h^{\nu b}: &\qquad {\mathrm i}\frac{2P^{2^+}_h-(1-12\alpha_E)P^{0^+}_{h_1}}{2p^2}\nonumber\\
  h^{\mu a}-\phi: &\qquad \frac{-\sqrt{6\alpha_E}{\mathrm i}P^{0^+}_{h_1\phi}}{p^2} \nonumber\\
  \phi-\phi: & \qquad \frac{{\mathrm i}}{p^2}
  \nonumber\\
  \Omega^{\mu ab}-\Omega^{\nu cd}: &-{\mathrm i} \frac{P^{0^-}_{\Omega}}{(2g_1+ g_2)p^2} - {\mathrm i}\frac{3P^{1^+}_{\Omega_1}}{(2g_1-2g_2+2g_4)p^2} -{\mathrm i}\frac{2P^{1^-}_{\Omega_1}}{(g_1-g_2 + 2g_4)p^2}- {\mathrm i}\frac{4P^{2^-}_{\Omega}}{(g_1-g_2)p^2}
\end{align}
 \paragraph{}
 
In general, when treating the fields $\chi_{\mu}{}^a$ and $\Omega_\mu{}^{ab}$ as Yang-Mills gauge fields in GQFT, we can simply add the usual gauge-fixing terms for the gauge-type gravifield $\chi_{\mu}{}^a$ and the spin gauge field $\Omega_\mu{}^{ab}$. For simplicity, we take the following explicit forms for their gauge fixing conditions 
\begin{align*}
& \lambda \hat{\chi}^{\mu\mu'}\hat{\chi}^{\nu\nu'}(\partial_\mu\chi_{\mu'}{}^a)(\partial_\nu\chi_{\nu' a})\, ;\quad \hat{\lambda}(\partial_\mu\hat{\chi}_a{}^\mu)(\partial_\nu\hat{\chi}^{a\nu}) ;\quad\xi \hat{\chi}^{\mu\mu'}\hat{\chi}^{\nu\nu'} (\partial_\mu\Omega_{\mu'}{}^{ab})(\partial_\nu\Omega_{\nu' ab});  \\
 & - \tfrac{\hat{\xi}}{4}\hat{\chi}^{\mu\rho} (\hat{\chi}^{b\nu}\partial_\nu\Omega_{\mu ab}+\chi_{\mu}{}^{a'}\hat{\chi}^{b\nu}\hat{\chi}_a{}^{\mu'}\partial_\nu\Omega_{\mu' a'b})(\hat{\chi}_c{}^{\sigma}\partial_\sigma\Omega_{\rho}{}^{ac}+\chi_{\rho d'}\hat{\chi}^{a\rho'}\hat{\chi}_c{}^{\sigma}\partial_\sigma\Omega_{\rho'}{}^{d'c}).
\end{align*}
In such a case, the coefficient matrices of the field equations are given by,
\begin{align}
a^{0^-}_{f_1 f_2} & =-(2g_1+g_2)p^2,\quad f_1= f_2 =\Omega \label{new0minus}
 \\
\nonumber \\
a^{0^+}_{f_1 f_2} & =\left(
\begin{array}{cccc}
-2\hat{\xi}p^2& 0& 0&  0\\
0& -2p^2 & 0& -2\sqrt{6\alpha_E}p^2\\
0& 0& (2\lambda+2\hat{\lambda})p^2& 0\\
0& -2\sqrt{6\alpha_E}p^2& 0& 1-12\alpha_Ep^2\\
\end{array} \right), \quad  f_1, f_2 =(\Omega,h_1,h_2,\phi)\label{new0plus}
\\
\nonumber \\
a^{1^-}_{f_1 f_2} &=\left(
\begin{array}{cccc}
\tfrac{g_2-g_1-2g_4}{2}p^2& 0& 0& 0\\
0& - (\hat{\xi} +2\xi) p^2& 0& 0\\
0& 0& (\lambda+\hat\lambda)p^2& (\lambda-\hat{\lambda})p^2\\
0& 0& (\lambda-\hat{\lambda})p^2& (\lambda+\hat{\lambda})p^2\\
\end{array} \right), \quad f_1, f_2 =(\Omega_1,\Omega_2,h_1,h_2) \label{new1minus}
\\
\nonumber \\
a^{1^+}_{f_1 f_2} &=\left(
\begin{array}{ccc}
\tfrac{2g_2-2g_1-2g_4-2\xi}{3}p^2& -\tfrac{\sqrt{2}(g_2-g_1-g_4+2\xi)}{3}p^2 &0\\
-\tfrac{\sqrt{2}(g_2-g_1-g_4+2\xi)}{3}p^2& \tfrac{g_2-g_1-g_4-4\xi}{3}p^2 &0\\
0 & 0 & 0\\
\end{array}\right),  \quad f_1, f_2=(\Omega_1,\Omega_2,h)\label{new1plus}
\\
\nonumber \\
a^{2^-}_{f_1 f_2} & =  \frac{g_2-g_1}{4}p^2, \quad f_1= f_2 =\Omega \label{new2minus}
\\
\nonumber \\
a^{2^+}_{f_1 f_2} & = \left(
\begin{array}{cc}
-2\hat{\xi}p^2& 0\\
0& p^2\\
\end{array}\right), \quad f_1 , f_2 =(\Omega,h) \label{new2plus}
\end{align}
Except for the $1^+$ component of the gravifield, all other coefficient matrices are non-degenerate. Thus we are able to inverse the matrices by requiring 
\begin{equation}
P^{1^+}_h h=0
\end{equation}
and get the propagators:
\begin{align}\label{new_prop}
  h^{\mu a}-h^{\nu b}: & {\mathrm i}\frac{2P^{2^+}_h-(1-12\alpha_E)P^{0^+}_{h_1}}{2p^2}  +{\mathrm i}\frac{P^{0+}_{h_2}}{2(\lambda+\hat{\lambda})p^2}\nonumber\\&+ {\mathrm i}\frac{(\lambda+\hat{\lambda})(P^{1^-}_{h_1}+P^{1^-}_{h_2})
  +(\lambda-\hat{\lambda})(P^{1^-}_{h_1h_2}+P^{1^-}_{h_2h_1})}{4\lambda\hat{\lambda}p^2} \nonumber\\
  h^{\mu a}-\phi: &\qquad \frac{-\sqrt{6\alpha_E}{\mathrm i}P^{0^+}_{h_1\phi}}{p^2} \nonumber\\
  \phi-\phi: & \qquad \frac{{\mathrm i}}{p^2} 
  \nonumber\\
  \Omega^{\mu ab}-\Omega^{\nu cd}: &-{\mathrm i} \frac{P^{0^-}_{\Omega}}{(2g_1+g_2)p^2} - {\mathrm i}\frac{P^{0^+}_{\Omega}}{2\hat{\xi}p^2} - i\frac{2P^{1^-}_{\Omega_1}}{(g_1-g_2+2g_4)p^2}- {\mathrm i}\frac{P^{1^-}_{\Omega_2}}{(2\xi +\hat{\xi})p^2} - {\mathrm i}\frac{4P^{2^-}_{\Omega}}{(g_1-g_2)p^2}
  \nonumber\\
  &- {\mathrm i}\frac{P^{2^+}_{\Omega}}{2\hat{\xi}p^2} -{\mathrm i}\frac{(4\xi +g_1-g_2+g_4)P^{1^+}_{\Omega_1}+(2\xi +2g_1-2g_2+2g_4)P^{1^+}_{\Omega_2}}{6\xi(g_1-g_2+g_4)p^2}
  \nonumber\\
  &+{\mathrm i} \frac{\sqrt{2}(2\xi-g_1+g_2-g_4)P^{1^+}_{\Omega_1\Omega_2}+\sqrt{2}(2\xi-g_1+g_2-g_4)P^{1^+}_{\Omega_2\Omega_1}}{6\xi(g_1-g_2+g_4)p^2} .
\end{align}
Taking $\xi,\hat{\xi},\lambda,\hat{\lambda}\rightarrow \infty$ as like the Landau gauge, the propagators are reduced to
\begin{align}\label{landau_prop}
  h^{\mu a}-h^{\nu b}: \quad & {\mathrm i}\frac{2P^{2^+}_h-(1-12\alpha_E)P^{0^+}_{h_1}}{2p^2}  \nonumber\\
  h^{\mu a}-\phi: &\qquad \frac{-\sqrt{6\alpha_E}{\mathrm i}P^{0^+}_{h_1\phi}}{p^2} \nonumber\\
  \phi-\phi: & \qquad \frac{{\mathrm i}}{p^2} 
  \nonumber\\
  \Omega^{\mu ab}-\Omega^{\nu cd}: \quad &-{\mathrm i} \frac{P^{0^-}_{\Omega}}{(2g_1+g_2)p^2} - {\mathrm i}\frac{2P^{1^-}_{\Omega_1}}{(g_1-g_2+2g_4)p^2}- {\mathrm i}\frac{4P^{2^-}_{\Omega}}{(g_1-g_2)p^2}
  \nonumber\\
  &-{\mathrm i}\frac{2 P^{1^+}_{\Omega_1}+P^{1^+}_{\Omega_2}-\sqrt{2}(P^{1^+}_{\Omega_1\Omega_2}+P^{1^+}_{\Omega_2\Omega_1})}{3(g_1-g_2+g_4)p^2}
\end{align}
It is seen that in this case the propagator of the gravifield recovers the same one as the case without adding gauge fixing condition, while the propagator of the spin gauge field is modified for the spin 1 component with even parity, which is relevant to the total antisymmetric part of spin gauge field.     

\paragraph{}
It is noticed that there is an intersection term $\phi-h$ which is caused as the choice of $h$ and $\phi$ is not orthogonal. To avoid such a complication, it is useful to redefine the quantum field
\begin{equation}
 h_{\mu a}\to H_{\mu a}=h_{\mu a}+\sqrt{2\alpha_E}\eta_{\mu a}\phi
 \end{equation}
so that the propagator of the field $H_{\mu a} $ becomes 
\begin{equation}
H^{\mu a}-H^{\nu b}:  {\mathrm i}\frac{\theta^{\mu\nu}\theta^{ab}+\theta^{\mu b}\theta^{\nu a}-\theta^{\mu a}\theta^{\nu b}}{2p^2} 
\end{equation}
which is compatible with the propagator in the usual linear gravity approach~\cite{Choi:1994ax} up to a gauge term $\frac{p^\mu p^\nu}{p^2}$.
If we take the gauge coefficients $\lambda,\hat{\lambda}$ to be $\frac{3}{2}$,
the explicit form of the $H-H$ propagator is
\begin{equation}\label{fixed_h_prop}
{\mathrm i}\frac{\eta^{\mu\nu}\eta^{ab}+\eta^{\mu b}\eta^{a\nu}-\eta^{\mu a}\eta^{\nu b}-\eta^{\mu\nu}\frac{p^ap^b}{3p^2}-\frac{p^\mu p^\nu}{3p^2}\eta^{ab}-\frac{p^\mu p^b}{p^2}\eta^{a\nu}-\eta^{\mu b}\frac{p^a p^\nu}{p^2}+\eta^{\mu a}\frac{p^\nu p^b}{p^2}+\frac{p^\mu p^a}{p^2}\eta^{\nu b}}{2p^2}
\end{equation}
When taking the gauge fixing parameters as follows 
\begin{equation}
\xi\to\infty,\quad \hat{\xi}^{-1}= \frac{8}{g_1-g_2}+\frac{4}{g_1-g_2+2g_4} -\frac{2}{g_1-g_2+g_4}
\end{equation}
the explicit form of the $\Omega-\Omega$ propagator is
\begin{align}\label{fixed_o_prop}
-&{\mathrm i}(\frac{4}{3 ( g_1 - g_2)} +\frac{1}{6 (2 g_1 + g_2)}) \frac{1}{p^2} (\eta^{\mu\nu}-\frac{p^\mu p^\nu}{p^2})( \eta^{a c}\eta^{bd}-\eta^{b c}\eta^{ad}) \nonumber\\
+ &  {\mathrm i} (\frac{1}{3 (g_1 -  g_2)} + \frac{1}{6 (2 g_1 + g_2)} -  \frac{1}{2 (g_1 -  g_2 + 2 g_4)})\frac{1}{p^4} \eta^{\mu\nu}(\eta^{ac}p^bp^d-\eta^{ad}p^bp^c-\eta^{bc}p^ap^d+\eta^{bd}p^ap^c)
\nonumber\\
-&{\mathrm i}(\frac{2}{3 ( g_1 - g_2)} - \frac{1}{6 (2 g_1 + g_2)})\frac{1}{p^2} (\eta^{\mu c}\eta^{bd}\eta^{\nu a}-\eta^{\mu d}\eta^{bc}\eta^{\nu a}-\eta^{\mu c}\eta^{ad}\eta^{\nu b}+\eta^{\mu d}\eta^{ac}\eta^{\nu b}-\frac{p^\mu p^c}{p^2}\eta^{bd}\eta^{\nu a}
\nonumber\\
&+\frac{p^\mu p^d}{p^2}\eta^{bc}\eta^{\nu a}+\frac{p^\mu p^c}{p^2}\eta^{ad}\eta^{\nu b}-\frac{p^\mu p^d}{p^2}\eta^{ac}\eta^{\nu b}-\frac{p^\nu p^a}{p^2}\eta^{bd}\eta^{\mu c}+\frac{p^\nu p^b}{p^2}\eta^{ad}\eta^{\mu c}+\frac{p^\nu p^a}{p^2}\eta^{bc}\eta^{\mu b}-\frac{p^\nu p^b}{p^2}\eta^{ac}\eta^{\mu d})
\nonumber\\
-&{\mathrm i}( \frac{1}{3 (g_1 -  g_2)} + \frac{1}{6 (2 g_1 + g_2)} -  \frac{1}{2 (g_1 - g_2 + g_4)} + \frac{1}{2 (g_1 -  g_2 + 2 g_4)})
\nonumber\\
& \frac{1}{p^4} (\eta^{\mu c}\eta^{a\nu}p^bp^d-\eta^{\mu d}\eta^{a\nu}p^bp^c-\eta^{\mu c}\eta^{b\nu}p^ap^d+\eta^{\mu d}\eta^{b\nu}p^ap^c) \nonumber \\
+&{\mathrm i}(\frac{1}{(g_1 -  g_2)} -  \frac{1}{2 (g_1 -  g_2 + 2 g_4)})\frac{1}{p^2}  (\eta^{\mu a}\eta^{bd}\eta^{\nu c}-\eta^{\mu b}\eta^{ad}\eta^{\nu c}-\eta^{\mu a}\eta^{bc}\eta^{\nu d}+\eta^{\mu b}\eta^{ac}\eta^{\nu d}-\eta^{\mu a}\eta^{bd}\frac{p^\nu p^c}{p^2}
\nonumber\\
&+\eta^{\mu b}\eta^{ad}\frac{p^\nu p^c}{p^2}+\eta^{\mu a}\eta^{bc}\frac{p^\nu p^d}{p^2}-\eta^{\mu b}\eta^{ac}\frac{p^\nu p^d}{p^2}-\eta^{\nu c}\eta^{bd}\frac{p^\mu p^a}{p^2}+\eta^{\nu d}\eta^{bc}\frac{p^\mu p^a}{p^2}+\eta^{\nu c}\eta^{ad}\frac{p^\mu p^b}{p^2}-\eta^{\nu d}\eta^{ac}\frac{p^\mu p^b}{p^2})
\nonumber\\
-&{\mathrm i}(\frac{1}{(g_1 -  g_2)} - \frac{1}{2 (g_1 -  g_2 + 2 g_4)})\frac{1}{p^4} (\eta^{\mu a}\eta^{\nu c}p^bp^d-\eta^{\mu b}\eta^{\nu c}p^ap^d-\eta^{\mu a}\eta^{\nu d}p^bp^c+\eta^{\mu b}\eta^{\nu d}p^ap^c), 
\end{align}
so that the highest order pole in the propagator is $\sim\frac{p^\mu p^\nu}{p^4}$ term, which behaves like a Yang-Mills gauge field propagator.
 In the following section, we will use the redefined symmetric quantum gravifield to calculate the physical observable.

\section{Gravitational scattering amplitude of Dirac spinor and modified Newton's law with background field }
\begin{figure}[h]
  \centering
  \begin{minipage}[h]{0.3\linewidth}
  \centering
  \includegraphics[width=0.8\linewidth]{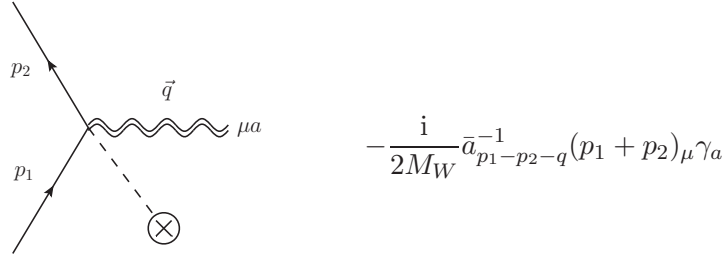}\\
  \end{minipage}
  \begin{minipage}[h]{0.3\linewidth}
  \begin{equation*}
  -\frac{{\mathrm i}}{2M_W}\bar{a}^{-1}_{p_1-p_2-q}(p_1+p_2)_\mu\gamma_a
  \end{equation*}
  \end{minipage}
  \caption{3-vertex for $f-f-h$.}\label{ffh_v}
\end{figure}

\begin{figure}[h]
  \centering
  \begin{minipage}[h]{0.3\linewidth}
  \centering
  \includegraphics[width=0.8\linewidth]{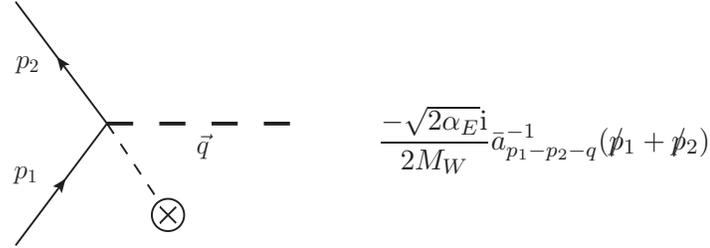}\\
  \end{minipage}
  \begin{minipage}[h]{0.3\linewidth}
  \begin{equation*}
  \frac{-\sqrt{2\alpha_E}{\mathrm i}}{2M_W}\bar{a}^{-1}_{p_1-p_2-q}(p\sslash_1+p\sslash_2)
  \end{equation*}
  \end{minipage}
  \caption{3-vertex for $f-f-\phi$.}\label{ffphi_v}
\end{figure}

\begin{figure}[h]
  \centering
  \includegraphics[width=0.5\linewidth]{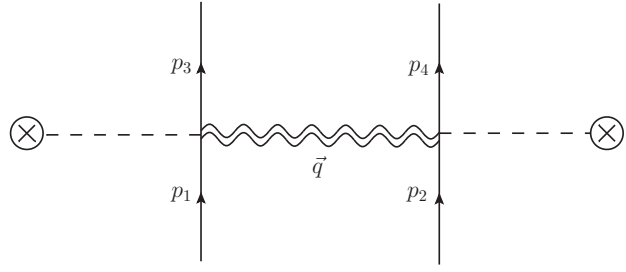}
  \caption{Tree diagram for 2-fermion scattering via gravifield.}\label{ffh_m}
\end{figure}

Let us now focus on the gravitational interaction between the Dirac spinor field $\Psi$ in the early universe. The leading order vertex of the fermion involves the background field.
\begin{equation}\label{vertex}
 - \frac{{\mathrm i}}{2\bar{a}(x)M_W}(H^{\mu a}-\sqrt{2\alpha_E}\phi\eta^{\mu a})\bar{\Psi}\partial_\mu\gamma_a\Psi + h.c.
  +\frac{1}{4}\Omega_{\mu ab}\bar{\Psi}\epsilon^{\mu ab\nu}\gamma_\nu\gamma_5\Psi \, .
\end{equation}
In the momentum space, the background scaling factor is given by  
\begin{equation}
\bar{a}^{-1}_k=\int d^4 x \frac{M_s\sqrt{\lambda_s/6\alpha_E}}{\sqrt{\kappa_\mu\kappa^\mu}}(1-\kappa\cdot x) {\mathrm e}^{{\mathrm i}x\cdot k}\nonumber \\=\frac{M_s\sqrt{\lambda_s/6\alpha_E}}{\sqrt{\kappa_\mu\kappa^\mu}}(\delta^4(k)+{\mathrm i}\kappa^\mu\partial_{k_\mu}\delta^{4}(k))
\end{equation}
where $\partial_{k_\mu}\equiv\frac{\partial}{\partial k_\mu}$. Corresponding to Feynman rules shown in Fig.\ref{ffh_v} and Fig.\ref{ffphi_v}.
\paragraph{}
Note that in calculating the fermion-fermion scattering, the gamma matrix in the vertex is contracted with the two external spinors, which satisfies the equation $\bar{u}_p p\sslash=p\sslash u_p=0$. So that the couplings to $\phi$ do not contribute to the tree-level diagrams. For the same reason, the third term from the $H$ propagator does not contribute to the result, either.
\paragraph{}
The tree level amplitude of the two-fermion scattering, with in-state momenta $p_1$ and $p_2$, and out-state momenta $p_3$ and $p_4$, is shown in Fig.~\ref{ffh_m}
\begin{align}
\label{amplitude}
<p_1p_2|{\mathrm i}T|p_3p_4>&=-\frac{{\mathrm i}\lambda_s}{96\alpha_E^2\kappa_\mu\kappa^\mu}\int \frac{dq^4}{(2\pi)^4}(\bar{u}_3^{s'}\gamma_\mu u_1^s \bar{u}_4^{r'}\gamma^\mu u_2^r(p_1+p_3)\cdot(p_2+p_4)+\bar{u}_3^{s'}(p\sslash_2+p\sslash_4) u_1^s \nonumber
\\& \bar{u}_4^{r'}(p\sslash_1+p\sslash_3) u_2^r) \frac{1}{q^2} \cdot [ \delta^4(p_1-p_3-q)-{\mathrm i}\kappa^\mu\partial_{q_\mu}\delta^{4}(p_1-p_3-q) ]\nonumber
\\& [ \delta^4(p_2-p_4+q)+{\mathrm i}\kappa^\mu\partial_{q_\mu}\delta^{4}(p_2-p_4+q) ] \, .
\end{align}
The main purpose is to check the newtonian potential in the early universe with the existence of background field. For the case that all fields are massless, we cannot take a non-relativistic limit to simplify the amplitude. Let us first check the cross section of this scattering process to contract all the spinors. After integrating the momenta of the propagator, the amplitude in \eqref{amplitude} becomes:
\begin{align}
\label{intamplitude}
<{\mathrm i}T>&=-\frac{{\mathrm i}\lambda_s}{96\alpha_E^2\kappa_\mu\kappa^\mu}(\bar{u}_3^{s'}\gamma_\mu u_1^s \bar{u}_4^{r'}\gamma^\mu u_2^r(p_1+p_3)\cdot(p_2+p_4)+\bar{u}_3^{s'}(p\sslash_2+p\sslash_4) u_1^s \bar{u}_4^{r'}(p\sslash_1+p\sslash_3) u_2^r) \nonumber
\\&\cdot \frac{1}{q^2}[ 1 + {\mathrm i} \frac{\kappa^\mu q_\mu}{q^2} - {\mathrm i}\kappa^\mu \partial_{ q_\mu} ] [ \delta^4(p_2-p_4+q)+{\mathrm i}\kappa^\mu\partial_{q_\mu}\delta^{4}(p_2-p_4+q) \, ] |_{q=p_1-p_3}
\end{align}
The derivatives of $\delta(p)$ can be expressed as some functions multiplied by $\delta(p)$, thus we can write the second line in \eqref{intamplitude} to the following general form
\begin{equation*}
 \delta^4(p_2-p_4+ p_1-p_3) \, \frac{1}{(p_1 - p_3)^2}[ 1 + F(p_1-p_3) ] \, .
\end{equation*}
Then our result of the scattering amplitude, except for the overall coefficient and $F(p_1-p_3)$ term that are related to the background field, is consistent with the leading order result shown in ~\cite{Ganjali:2018hyj}. If we were working in another gauge fixing condition, the difference would be terms proportional to $\frac{q^\mu q^\nu}{q^2}$, contracted with the vertex will gives the term $\bar{u}_3 (p\sslash_3-p\sslash_1)u_1$ or $(p_3-p_1)\cdot(p_3+p_1)$, both of which are vanishing because of the on-shell condition of the external fermions. So our result is indeed gauge independent.
\paragraph{}
The squared matrix element, after throwing all the spin information, is:
\begin{align}\label{squaredm}
 \frac{1}{4}\sum_{r,r',s,s'}|\mathcal{M}|^2&=(\frac{\lambda_s}{192\alpha_E^2\kappa_\mu\kappa^\mu})^2 \frac{1}{(p_1 - p_3)^4} |1 + F(p_1-p_3)|^2\nonumber
 \\&\cdot \sum_{r,r',s,s'}((p_1+p_3)\cdot(p_2+p_4))^2\Tr[u_2^{s'}\bar{u}_2^{s'}\gamma^\nu u_4^{r'}\bar{u}_4^{r'}\gamma_\rho]\Tr[u_1^{s}\bar{u}_1^{s}\gamma_\nu u_3^{r}\bar{u}_3^{r}\gamma^\rho]\nonumber
 \\ &+ (p_1+p_3)\cdot(p_2+p_4)\Tr[u_4^{r'}\bar{u}_4^{r'}(p\sslash_4+p\sslash_2) u_2^{s'}\bar{u}_2^{s'}\gamma_\rho]\Tr[u_3^{r}\bar{u}_3^{r} (p\sslash_3+p\sslash_1) u_1^{s}\bar{u}_1^{s}\gamma^\rho]\nonumber
 \\ &+ (p_1+p_3)\cdot(p_2+p_4)\Tr[u_4^{r'}\bar{u}_4^{r'}\gamma_\rho u_2^{s'}\bar{u}_2^{s'}(p\sslash_4+p\sslash_2) ]\Tr[u_3^{r}\bar{u}_3^{r} \gamma^\rho u_1^{s}\bar{u}_1^{s} (p\sslash_3+p\sslash_1)]\nonumber
 \\ &+ \Tr[u_4^{r'}\bar{u}_4^{r'}(p\sslash_1+p\sslash_3) u_2^{s'}\bar{u}_2^{s'}(p\sslash_1+p\sslash_3)] \Tr[u_3^{r}\bar{u}_3^{r} (p\sslash_2+p\sslash_4) u_1^{s}\bar{u}_1^{s} (p\sslash_2+p\sslash_4)]\, .
\end{align}
With the spin sum rules $\sum_s u^s(p)\bar{u}^s(p)=p\sslash$, and the Mandelstam variables~\cite{Peskin:1995ev}
\begin{align*}
  s&=(p_1+p_2)^2=(p_3+p_4)^2=2p_1\cdot p_2=2p_3\cdot p_4\\
  t&=(p_1-p_3)^2=(p_2-p_4)^2=-2p_1\cdot p_3=-2p_2\cdot p_4\\
  u&=(p_1-p_4)^2=(p_3-p_2)^2=-2p_1\cdot p_4=-2p_3\cdot p_2\\
  0&=s+t+u;
\end{align*}
we can simplify \eqref{squaredm} into the follow form
\begin{align}\label{symsquaredm}
 \frac{1}{4}\sum_{r,r',s,s'}|\mathcal{M}|^2&=(\frac{\lambda_s}{192\alpha_E^2\kappa_\mu\kappa^\mu})^2  \frac{1}{t^2} |1 + F(p_1-p_3) |^2\nonumber
 \\ &\cdot[\, 8(s^2+u^2)(s-u)^2-16(s-u)^2su-16(s-u)^2su+64s^2u^2 \, ]
\end{align}
As long as the two massless fermions are not in the same direction, we can always make a Lorentz boost to a center-of-energy frame, so that  $\vec{p}_1=-\vec{p}_2$ and $p_1^0=p_2^0=E$. When taking the weak interaction limit that $\theta \to 0$, we have
\begin{align}\label{crosssection}
 s&=4E^2;\quad t= - 2E^2(1-cos\theta);\qquad u=- 2E^2(1+cos\theta);\nonumber\\
 \frac{d\sigma}{d\Omega}&=\frac{|\mathcal{M}|^2}{64\pi^2 (2E)^2} =(\frac{\lambda_s}{192\alpha_E^2\kappa_\mu\kappa^\mu})^2 \frac{1}{(1-cos\theta)^2} | 1 + F(p_1-p_3) \, |^2\nonumber
 \\ &\quad\cdot \frac{E^2}{8\pi^2}(149+232cos\theta+114cos^2\theta+16cos^3\theta+cos^4\theta)\nonumber
 \\ &\xrightarrow{\theta\rightarrow0}(\frac{E^3\lambda_s}{12\pi\alpha_E^2\kappa_\mu\kappa^\mu})^2\, \frac{1}{(\vec{p}_1-\vec{p}_3)^4} |1  + F(p_1-p_3)|^2 \, .
\end{align}
In comparison with the Born approximation of the cross section~\cite{Schwartz:2013pla} 
\begin{align}
(\frac{d\sigma}{d\Omega})_{Born}=\frac{E^2}{4\pi^2}|\tilde{V}(\vec{q})|^2,
\end{align} 
To compare our result with those from the usual Newtonian potential, we identify the factor $\frac{1}{M_W^2}$ with the coefficient of the Einstein equation $8\pi G$. So the relation between $\alpha_E$ and Newtonian gravitational constant $G_N$is
\begin{equation}
\alpha_E=\frac{M_w^2}{2M_s^2}=\frac{1}{16\pi G_N M_s^2}
\end{equation}
 Then we we obtain the potential in the momentum space as:
\begin{align}\label{potential}
  \tilde{V}(\vec{q})&= - \frac{\lambda_sM_s^2}{6\alpha_E\kappa_\mu\kappa^\mu}\, \frac{16\pi G_N E^2}{\vec{q}^{\; 2}}[ 1 + F(\vec{q})]
\end{align}
The leading term will contribute to a $1/r$ potential in the coordinate space.  Such a term coincides with the Newton's law, but it is modified by a factor $\tfrac{\lambda_sM_s^2}{6\alpha_E\kappa^2}$ which depends on the size of the inverse of scaling factor $1/\bar{a}(x)$. In the early universe, the scaling factor is much smaller, thus the gravitational potential can become much stronger. The modified term $F(\vec{q})$ contains the structure of the derivatives of delta functions, we shall investigate its effect elsewhere. 
\paragraph{}
Note that the coefficient $16\pi G_N$  is four times than the gravitational potential for the massive Dirac fermions. This is because we are working on the massless Dirac fermions. When considering the Dirac fermion getting a mass from spontaneous symmetry breaking, a mass term will be generated. In a unitary scaling gauge condition $\det{\chi}=1$, we need to consider the change of the spinor structure, and an additional 
\begin{equation}\label{addtional_mass}
-\bar{u}_3^{r'}(p\sslash_3+p\sslash_1)u_1^r\bar{u}_4^{s'}(p\sslash_4+p\sslash_2)u_2^s=-4m^2\bar{u}^{r'}_3u_1^r\bar{u}_4^{s'}u_2^s=-16m^4\delta^{r'r}\delta^{s's}
\end{equation}
from the third term~\eqref{fixed_h_prop} of the graviton propagator. The massive Dirac fermion allows us to take a non-relativistic approximation
\begin{align}
u_p=\begin{pmatrix}
\sqrt{p^\mu\sigma_\mu}\xi
\\
\sqrt{p^\mu\bar{\sigma}_\mu}\xi
\end{pmatrix}
=
\begin{pmatrix}
\sqrt{E-\vec{p}\cdot\vec{\sigma}}\xi
\\
\sqrt{E+\vec{p}\cdot\vec{\sigma}}\xi
\end{pmatrix}
\xrightarrow{\frac{\vec{p}}{E}\rightarrow0}\sqrt{E}
\begin{pmatrix}
(1-\frac{1}{2E}\vec{p}\cdot\vec{\sigma})\xi
\\
(1+\frac{1}{2E}\vec{p}\cdot\vec{\sigma})\xi
\end{pmatrix}+O(\frac{p^2}{E^2}) .
\end{align}
The leading order and next-to-leading-order contributions from $\bar{u}_{p_3}\gamma_{\mu}u_{p_1}$ is found to be
\begin{align}
\bar{u}_{p_3}\gamma_{\mu}u_{p_1}\xrightarrow{\frac{\vec{p}}{E}\rightarrow0}&\sqrt{E_3E_1}\begin{pmatrix}
\xi^{r\dagger}(1-\frac{1}{2E_3}\vec{p_3}\cdot\vec{\sigma}), &\xi^{r\dagger}(1+\frac{1}{2E_3}\vec{p_3}\cdot\vec{\sigma})
\end{pmatrix}
\begin{pmatrix}
\sigma^\mu\\
&\bar{\sigma}^\mu
\end{pmatrix}
\begin{pmatrix}
(1-\frac{1}{2E_1}\vec{p_1}\cdot\vec{\sigma})\xi^{r'}
\\
(1+\frac{1}{2E_1}\vec{p_1}\cdot\vec{\sigma})\xi^{r'}
\end{pmatrix}
\nonumber\\
=&\xi^{r\dagger}(2\sqrt{E_3E_1}\eta^{0\mu}-g^{i\mu}\vec{p}_3\cdot\vec{\sigma}\sigma^i-g^{i\mu}\sigma^i\vec{p}_1\cdot\vec{\sigma})\xi^{r'}+O(\frac{p^2}{E^2})
\end{align}
The leading term for $\mu=0$ requires $r=r'$, which together with \eqref{addtional_mass} enables us to get a factor $\frac{1}{4}$ for the potential~\eqref{potential}. The next-to-leading-order term for $\mu=0$ comes from the expansion of $E$
\begin{align}
4\sqrt{E_1E_2E_3E_4}&=4m^2\sqrt[4]{(1+\frac{\vec{p}_1^2}{m^2})(1+\frac{\vec{p}_2^2}{m^2})(1+\frac{\vec{p}_3^2}{m^2})(1+\frac{\vec{p}_4^2}{m^2})}=4m^2+\vec{p}_1^2+\vec{p}_2^2+\vec{p}_3^2+\vec{p}_4^2
\nonumber\\
&=2\sqrt{E_1E_3}(E_2+E_4)=2\sqrt{E_2E_4}(E_1+E_3)
\end{align}

\paragraph{}
The next-to-leading order from $\mu=i$ can be simplified to
\begin{align}
&-\xi^{r\dagger}(\frac{(\vec{p}_3+\vec{p}_1)_j}{2}\{\sigma^j,\sigma^i\}+\frac{(\vec{p}_3-\vec{p}_1)_j}{2}\cdot[\sigma^j,\sigma^i])\xi^{r'}
\nonumber\\=&i\xi^{r\dagger}\epsilon^{ijk}(\vec{p}_3-\vec{p}_1)_j\sigma_k\xi^{r'}-\xi^{r\dagger}\xi^{r'}(\vec{p}_3+\vec{p}_1)^i \, ,
\end{align}
the spinor formalism can be re-expressed as a four-vector
\begin{align}
\bar{u}_{p_3}\gamma_{\mu}u_{p_1}&=(2m+\frac{\vec{p}_3^2+\vec{p}_1^2}{2m},-(\vec{p}_3+\vec{p}_1)+i(\vec{p}_3-\vec{p}_1)\times \vec{S_1}) ,\quad S_1^i\equiv\xi^\dagger\sigma^i\xi.
\end{align}
Substituting it into the expression  of the amplitude Eq.~\eqref{amplitude}
\begin{equation}
(\bar{u}_3^{s'}\gamma_\mu u_1^s \bar{u}_4^{r'}\gamma^\mu u_2^r(p_1+p_3)\cdot(p_2+p_4)+\bar{u}_3^{s'}(p\sslash_2+p\sslash_4) u_1^s  \bar{u}_4^{r'}(p\sslash_1+p\sslash_3) u_2^r)-16m^4\delta^{r'r}\delta^{s's} \, ,
\end{equation}
we can obtain the total contribution up to next-to-leading order,
\begin{align}
&8m^2\left(4m^2+(\vec{p}_1^2+\vec{p}_2^2+\vec{p}_3^2+\vec{p}_4^2)+( (\vec{p}_1 - \vec{p}_3)\times \vec{S}_1 ) \cdot ( (\vec{p}_2 - \vec{p}_4)\times \vec{S}_2) \right) -16m^4
\nonumber\\
=&16m^4\left(1+ \frac{1}{2m^2} \left( \sum_{i=1}^{4}\vec{p}_i^2 - \vec{q}^2\vec{S}_1\cdot\vec{S}_2+\vec{q}\cdot\vec{S}_2\vec{q}\cdot\vec{S}_1\right) \right).
\end{align}
So the potential for massive fermions is
\begin{align}\label{potential}
  \tilde{V}(\vec{q})&= \frac{-\lambda_sM_s^2}{6\alpha_E\kappa_\mu\kappa^\mu}\, \frac{16\pi G_N }{\vec{q}^{\; 2}}[ 1+ F(\vec{q})]\cdot\left(\frac{m^2}{4}+ \frac{1}{8} \left( \sum_{i=1}^{4}\vec{p}_i^2 - \vec{q}^2\vec{S}_1\cdot\vec{S}_2+\vec{q}\cdot\vec{S}_2\vec{q}\cdot\vec{S}_1)\right) \right)
\end{align}
Ignoring the kinematic energies, the next-to-leading order effect is proportional to the inner product of two particles, i.e.,  $-[ \vec{q}\times \vec{S}_1] \cdot [\vec{q}\times \vec{S}_2] $.
\paragraph{}
If we consider the anti-fermion, its spinor structure is
\begin{align}
v_p=\begin{pmatrix}
\sqrt{p^\mu\sigma_\mu}\eta
\\
-\sqrt{p^\mu\bar{\sigma}_\mu}\eta
\end{pmatrix}
=
\begin{pmatrix}
\sqrt{E-\vec{p}\cdot\vec{\sigma}}\eta
\\
-\sqrt{E+\vec{p}\cdot\vec{\sigma}}\eta
\end{pmatrix}
\xrightarrow{\frac{\vec{p}}{E}\rightarrow0}\sqrt{E}
\begin{pmatrix}
(1-\frac{1}{2E}\vec{p}\cdot\vec{\sigma})\eta
\\
-(1+\frac{1}{2E}\vec{p}\cdot\vec{\sigma})\eta
\end{pmatrix}+O(\frac{p^2}{E^2}) .
\end{align}
and the vertex would have a minus sign from $-(p_2+p_4)_\mu$. The vertex spinor contraction is
\begin{align}
\bar{v}_{p_2}\gamma_{\mu}v_{p_4}\xrightarrow{\frac{\vec{p}}{E}\rightarrow0}&\sqrt{E_2E_4}\begin{pmatrix}
\eta^{r\dagger}(1-\frac{1}{2E_2}\vec{p_3}\cdot\vec{\sigma}), &-\eta^{r\dagger}(1+\frac{1}{2E_2}\vec{p_3}\cdot\vec{\sigma})
\end{pmatrix}
\begin{pmatrix}
\sigma^\mu\\
&\bar{\sigma}^\mu
\end{pmatrix}
\begin{pmatrix}
(1-\frac{1}{2E_4}\vec{p_1}\cdot\vec{\sigma})\eta^{r'}
\\
-(1+\frac{1}{2E_4}\vec{p_1}\cdot\vec{\sigma})\eta^{r'}
\end{pmatrix}
\nonumber\\
=&\eta^{r\dagger}(2\sqrt{E_2E_4}g^{0\mu}-g^{i\mu}\vec{p}_3\cdot\vec{\sigma}\sigma^i-g^{i\mu}\sigma^i\vec{p}_1\cdot\vec{\sigma})\eta^{r'}+O(\frac{p^2}{E^2})
\end{align}
So the there was only an overall minus sign from the momentum, and will be compensated by the commutation of the fermion operator in the Wick contraction, thus the amplitude does not flip sign. The only possible difference lies in spin of the anti-fermion $\eta^\dagger \sigma^i\eta$. Thus we may use a separate spin notation to distinguish particle and anti-particle
\begin{equation}
\vec{S}^+\equiv\xi^\dagger \sigma^i\xi,\qquad \vec{S}^-\equiv\eta^\dagger \sigma^i\eta
\end{equation}
So the next-to-leading order effect between fermion and anti-fermion is 
\begin{equation}
-[ \vec{q}\times \vec{S}^+_1 ] \cdot  [ \vec{q}\times \vec{S}^-_2 ].
\end{equation}

\paragraph{}
Let us now consider the special case that the two massless ingoing particles are in the same direction. Suppose that their momenta are chosen as follows
\begin{equation}\label{samedir}
  p^\mu_1=(E_1,0,0,E_1),\qquad p^\mu_2=(E_2,0,0,E_2)\, .
\end{equation}
As the overall $\delta^4(p_1-p_3+p_2-p_4)$ guarantees the momentum conservation, the outgoing momenta must be in the same direction. In this case, all the momenta are in the same direction, they are null vectors. So that their product gives zero, namely $s=t=u=0$. As a consequence, the cross-section becomes vanishing.

\section{Scattering amplitude of the Dirac spinor via the spin gauge field}\label{sec:spin_gauge_int}

\begin{figure}[h]
  \centering
  \begin{minipage}[h]{0.3\linewidth}
  \centering
  \includegraphics[width=0.8\linewidth]{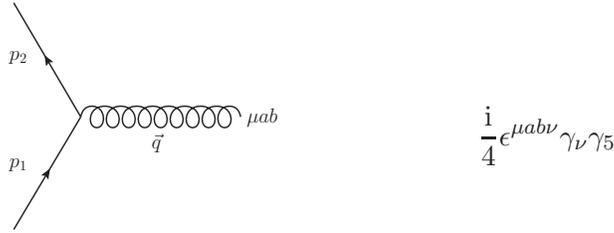}\\
  \end{minipage}
  \begin{minipage}[h]{0.3\linewidth}
  \begin{equation*}
  \frac{{\mathrm i}}{4}\epsilon^{\mu ab\nu}\gamma_\nu\gamma_5
  \end{equation*}
  \end{minipage}
  \caption{3-vertex for $f-f-\Omega$.}\label{ffs_v}
\end{figure}

It is interesting to consider the scattering amplitude of Dirac spinor via the spin gauge field. The leading order spin gauge interaction of Dirac spinor is given by the totally antisymmetric coupling of the spin gauge field. The vertex Feynman rule in Fig.\ref{ffs_v} can be derived from the last term in \eqref{vertex}. 

\begin{figure}[h]
  \centering
  \includegraphics[width=0.3\linewidth]{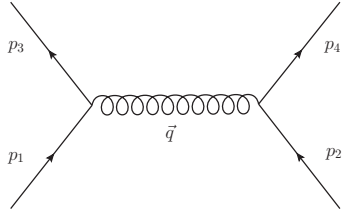}
  \caption{Tree diagram for 2-fermion scattering via spin gauge field.}\label{ffs_m}
\end{figure}

The propagator of the totally antisymmetric part of the spin gauge field is taken the following form
\begin{align}
\frac{-{\mathrm i}P^{0^-}_\Omega}{(2g_1+g_2)p^2} - \frac{2{\mathrm i}P^{1^+}_{\Omega_1}}{3(g_1-g_2+g_4)p^2}.
\end{align}
We may redefine the coupling constants~\cite{Wu:2017urh}
\begin{align*}
& g_1^{-1}  = g_h^2, \quad g_2/g_1 = \alpha_W,\quad  g_4/g_1 = \beta_W, \\
&  \frac{2}{3(g_1-g_2 +g_4)} =  \frac{2g_h^2}{3(1-\alpha_W +\beta_W)} , 
\end{align*}
and redefine the spin gauge field and replace the vertex
\begin{align*}
\Omega\rightarrow g_h \Omega,\qquad  \frac{{\mathrm i}g_h}{4}\epsilon^{\mu ab\nu}\gamma_\nu\gamma_5 .
\end{align*}
The Dirac spinor scattering amplitude via the spin gauge field is shown in Fig.\ref{ffs_m}
\begin{equation}\label{ffoamplitude}
  {\mathrm i}\mathcal{M}= \frac{{\mathrm i}g_h^2}{4(1-\alpha_W + \beta_W)}\frac{1}{q^2}\bar{u}_3^{s'}\gamma_\nu\gamma_5 u_1^s \bar{u}_4^{r'}\gamma^\nu\gamma_5 u_2^r .
\end{equation}
If the Dirac spinor acquires a mass from some symmetry breaking, we may take the non-relativistic limit of this amplitude. Different from the Coulomb potential where the leading contribution comes from $\bar{u}_3^{s'}\gamma_0u_1^s$~\cite{Peskin:1995ev}, the $\gamma_5$ in \eqref{ffoamplitude} will lead to 
\begin{align}
&u^s(p)\xrightarrow{\frac{\vec{p}^2}{m^2}\rightarrow 0}\sqrt{m}(\xi^s,\xi^s)^T,&\bar{u}_3^{s'}\gamma_0\gamma_5 u_1^s\xrightarrow{\frac{\vec{p}^2}{m^2}\rightarrow 0}m\xi^{s'\dagger}\xi^s-m\xi^{s'\dagger}\xi^s=0;\nonumber
\\&\bar{u}_3^{s'}\gamma_i\gamma_5 u_1^s\xrightarrow{\frac{\vec{p}^2}{m^2}\rightarrow 0}2m\xi^{s'\dagger}\sigma_i\xi^s, & \bar{u}_3^{s'}\gamma_\nu\gamma_5 u_1^s \bar{u}_4^{r'}\gamma^\nu\gamma_5 u_2^r\xrightarrow{\frac{\vec{p}^2}{m^2}\rightarrow 0}-\sum_i4m^2\xi^{s'\dagger}\sigma_i\xi^s \xi^{r'\dagger}\sigma_i\xi^r .
\end{align}
It is shown that the potential for 2-fermion scattering without spin change can be attractive (repulsive) for aligned spins and repulsive (attractive) for opposed spins, which relies on the sign of the coefficient $(1-\alpha_W + \beta_W)$ whether it is positive $1-\alpha_W + \beta_W >0$ (negative $1-\alpha_W + \beta_W < 0$). The potential of the totally antisymmetric field was studied in a different way in ref.~\cite{Kerlick:1975tr}, which arrived at the case of negative coefficient $1-\alpha_W + \beta_W<0$.  Such an interaction is independent of the background field. In the early universe, the scaling factor is so small that the gravitational effect becomes dominant to the cross sections.  The spin gauge coupling is no longer significant, its cross section is found to be:
\begin{equation}
\frac{d\sigma}{d\Omega}=\frac{(s^2+u^2)g_h^4}{8192\pi^2(1-\alpha_W +\beta_W)^2 E^2q^4}=\frac{(5+2cos\theta+cos^2\theta)g_h^4}{8192\pi^2(1-\alpha_W +\beta_W)^2E^2(1-cos\theta)^2} .
\end{equation}
When taking the weak interaction limit that $\theta \to 0$, we have
\begin{equation}
\frac{d\sigma}{d\Omega}=\frac{E^2g_h^4}{256\pi^2 (1-\alpha_W +\beta_W)^2 }\, \frac{1}{\vec{q}^{\;4}} ,
\end{equation}
which leads to a $1/r$ potential in the coordinate space.

\section{Conclusion}
We have investigated the gravitational interactions with the background field in the framework of GQFT. The full action of the GQFT with spin gauge and scaling gauge transformations has been expanded in a non-constant background field. To the leading order gravitational interactions in GQFT, we have derived the Feynman rules for the propagators and interacting vertices of the quantum fields by using the tensor projection operators. The quantum gravifield has been redefined to be normalized and diagonal, which leads to an interaction between the Dirac spinor and scalar fields. In the leading order,  the scalar interaction with the Dirac spinor vanishes when the massless Dirac spinor are on-mass shell as the external fields. We have calculated the tree-level two Dirac spinors scattering through the gravitational interaction and analyzed its amplitude and cross section. Besides the modified term from the derivative of delta function, the overall amplitude is proportional to the inverse of the scaling factors, which implies that the gravitational potential is much stronger in the early universe. The spin dependence of the gravitational potential in the nonrelativistic case has been analyzed.  We have also calculated the interaction between the Dirac spinor and the totally antisymmetric part of the spin gauge field at the leading order, which is similar to the result of the scattering through a vector field, but with a flip sign in the amplitude due to the property of axial vector, resulting in a spin gauge force which depends on the sign of the coefficient in its quadratic terms.

\section* {Acknowledgements}

This work was supported in part by the National Science Foundation of China
(NSFC) under Grant Nos. 11690022 and 11475237, and by the Strategic
Priority Research Program of the Chinese Academy of Sciences
(CAS), Grant No. XDB23030100, Key Research program QYZDYSSW-
SYS007, and by the CAS Center for Excellence in Particle Physics
(CCEPP).

\section*{Appendix A: Next-to-leading order quadratic terms}
We have presented the leading order quadratic terms in the context, the following are the higher order terms of the background field. We define
\begin{equation}
\hat{\kappa}^{\mu}\equiv\frac{\kappa^\mu}{\sqrt{\kappa^\nu\kappa_\nu}}.
\end{equation}
\paragraph{}
The next-to-leading order quadratic terms for $h_\mu{}^a-\phi$ are:
\begin{align*}
\phi&\left(- \sqrt{2} M_s^2 \lambda_s \hat{\kappa}^{a} \hat{\kappa}^{\mu} \bar{a}^2 -2\sqrt{\frac{\lambda_s}{3}}M_s \hat{\kappa}^{\mu} \bar{a}  \partial^{a}+2  \sqrt{\frac{\lambda_s}{3\alpha_E^2}} M_s\hat{\kappa}^{\mu}  \bar{a} \partial^{a}- 4\sqrt{\frac{\lambda_s}{3}}M_s \hat{\kappa}^{\nu} \eta^{\mu a}\bar{a} \partial_{\nu}\right)h_{\{\mu a\}}
\nonumber\\
&+ \phi \left(2\sqrt{3\lambda_s}M_s\hat{\kappa}^{\mu} \bar{a}  \partial^{a}\right) h_{[\mu a]}
\end{align*}
The terms for $h_\mu{}^a-w_\nu$ are:
\begin{equation*}
w_\nu\left(-2\sqrt{\frac{\lambda_s}{3\alpha_E^2}}g_wM_s^2\hat{\kappa}^a\eta^{\mu\nu}\bar{a}^2\right)h_{\{\mu a\}}
\end{equation*}
The terms for $\phi-w_\mu$ are:
\begin{equation*}
w_\mu\left(\sqrt{\frac{3\lambda_s}{2\alpha_E}}g_wM_s^2\hat{\kappa}^\mu\bar{a}^2+g_wM_s\bar{a}\partial^{\mu}\right)\phi
\end{equation*}
The terms for $w_\mu-w_\nu$ are:
\begin{equation*}
w_\mu\left(\frac{1}{2}g_w^2M_s^2\bar{a}^2\right)w_\nu
\end{equation*}
The terms for $\phi-\phi$ are:
\begin{equation*}
\phi\left(6\lambda_sM_s^2\bar{a}^2\right)\phi
\end{equation*}
The terms for $h_\mu{}^a-h_\nu{}^b$ are:
\begin{align*}
h_{\{\mu a\}}&\left(-\frac{\lambda_sM_s^2}{3\alpha_E}\eta^{\mu\nu}\eta^{ab}\bar{a}^2-\frac{\lambda_sM_s^2}{6\alpha_E}\eta^{\mu a}\eta^{\nu b}\bar{a}^2+\frac{\lambda_sM_s^2}{2\alpha_E^2}\hat{\kappa}^\mu\hat{\kappa}^\nu\eta^{a b}\bar{a}^2-\frac{\lambda_sM_s^2}{2\alpha_E}\hat{\kappa}^\mu\hat{\kappa}^\nu\eta^{a b}\bar{a}^2-\frac{\lambda_sM_s^2}{3\alpha_E}\hat{\kappa}^\mu\hat{\kappa}^a\eta^{\nu b}\bar{a}^2\right.
\nonumber\\
&+\left.\sqrt{\frac{8\lambda_s}{3\alpha_E}}M_s\hat{\kappa}^\mu\eta^{\nu b}\bar{a}\partial^{a}-\sqrt{\frac{2\lambda_s}{3\alpha_E}}M_s\hat{\kappa}^\mu\eta^{ab}\bar{a}\partial^\nu\right)h_{\{\nu b\}}+
\nonumber\\
h_{[\mu a]}&\left(-\frac{\lambda_sM_s^2}{3\alpha_E^2}\hat{\kappa}^{\mu}\hat{\kappa}^{\nu}\eta^{ab}\bar{a}^2+\sqrt{\frac{2\lambda_s}{3\alpha_E}}M_s\hat{\kappa}^\mu \eta^{\nu b}\bar{a}\partial^{a}\right)h_{\{\nu b\}}
+h_{[\mu a]}\left(\frac{\lambda_sM_s^2}{2\alpha_E}\eta^{\mu\nu}\eta^{ab}\bar{a}^2-\frac{\lambda_sM_s^2}{6\alpha_E^2}\hat{\kappa}^\mu\hat{\kappa}^\nu\eta^{ab}\bar{a}^2\right)h_{[\mu a]}
\end{align*}

\section*{Appendix B: Leading order vertices}

We have presented the leading order vertices of the fermions in the context, the following are the 3-vertices for the spin gauge field $\Omega_\mu{}^{ab}$ with the redefined field $\Omega$ by a coupling constant:
\begin{align*}
  &g_h^3\left(-(g_1+g_2) \Omega^{abc} \Omega^{d}{}_{b}{}^{m} \partial_{a}\Omega_{cdm} +  g_4 \Omega^{abc} \Omega_{bc}{}^{d} \partial_{a}\Omega^{m}{}_{dm} +  (g_1+g_2) \Omega^{abc} \Omega^{d}{}_{b}{}^{m} \partial_{c}\Omega_{adm} \right.
  \\-& 2 g_1 \Omega^{abc} \Omega^{d}{}_{b}{}^{m} \partial_{d}\Omega_{acm} +  (g_1+g_2) \Omega^{abc} \Omega^{d}{}_{b}{}^{m} \partial_{d}\Omega_{cam} - g_4 \Omega^{abc} \Omega_{bc}{}^{d} \partial_{d}\Omega^{m}{}_{am}- g_4 \Omega^{a}{}_{a}{}^{b} \Omega^{c}{}_{b}{}^{d} \partial_{c}\Omega^{m}{}_{dm} 
  \\+&  g_4 \Omega^{a}{}_{a}{}^{b} \Omega^{c}{}_{b}{}^{d} \partial_{d}\Omega^{m}{}_{cm} +  (g_1+g_2) \Omega^{abc} \Omega^{d}{}_{b}{}^{m} \partial_{m}\Omega_{acd} - g_4 \Omega^{abc} \Omega_{bc}{}^{d} \partial_{m}\Omega_{ad}{}^{m} - 2 g_2 \Omega^{abc} \Omega^{d}{}_{b}{}^{m} \partial_{m}\Omega_{cad}
  \\  +& \left.  g_4 \Omega^{a}{}_{a}{}^{b} \Omega^{c}{}_{b}{}^{d} \partial_{m}\Omega_{cd}{}^{m} +  g_4 \Omega^{abc} \Omega_{bc}{}^{d} \partial_{m}\Omega_{da}{}^{m} - g_4 \Omega^{a}{}_{a}{}^{b} \Omega^{c}{}_{b}{}^{d} \partial_{m}\Omega_{dc}{}^{m} \right) .
\end{align*}
For the gravifield $h_\mu{}^a$ interactions, we have 
\begin{align*}
  &\frac{1}{M_W \bar{a}(x)}\left(- \tfrac{1}{2} h^{ab} \partial_{a}h^{cd} \partial_{b}h_{cd} -  \tfrac{1}{2} h^{ab} \partial_{a}h^{cd} \partial_{b}h_{dc} + h^{ab} \partial_{a}h^{c}{}_{c} \partial_{b}h^{d}{}_{d} -  h^{ab} \partial_{b}h^{d}{}_{d} \partial_{c}h_{a}{}^{c} \right.
  \\-&  h^{ab} \partial_{a}h^{d}{}_{d} \partial_{c}h_{b}{}^{c} -  h^{ab} \partial_{b}h^{c}{}_{a} \partial_{c}h^{d}{}_{d} + h^{ab} \partial_{c}h^{d}{}_{d} \partial^{c}h_{ba} + h^{ab} \partial_{c}h_{a}{}^{c} \partial_{d}h_{b}{}^{d} 
  \\+& h^{ab} \partial_{b}h^{c}{}_{a} \partial_{d}h_{c}{}^{d} -  h^{a}{}_{b} \partial^{c}h^{b}{}_{a} \partial_{d}h_{c}{}^{d}  + \tfrac{1}{2} h^{ab} \partial_{b}h_{cd} \partial^{d}h_{a}{}^{c} + \tfrac{1}{2} h^{ab} \partial_{b}h_{dc} \partial^{d}h_{a}{}^{c}
  \\-&  \tfrac{1}{2} h^{ab} \partial_{c}h_{bd} \partial^{d}h_{a}{}^{c} -  \tfrac{1}{2} h^{ab} \partial_{d}h_{bc} \partial^{d}h_{a}{}^{c} + \tfrac{1}{2} h^{ab} \partial_{a}h_{cd} \partial^{d}h_{b}{}^{c} + \tfrac{1}{2} h^{ab} \partial_{a}h_{dc} \partial^{d}h_{b}{}^{c}
  \\ +&\left.  \tfrac{1}{2} h^{ab} \partial_{c}h_{da} \partial^{d}h_{b}{}^{c} -   \tfrac{1}{2} h^{ab} \partial_{d}h_{ca} \partial^{d}h_{b}{}^{c} -  \tfrac{1}{2} h^{ab} \partial_{b}h_{cd} \partial^{d}h^{c}{}_{a} + \tfrac{1}{2} h^{ab} \partial_{b}h_{dc} \partial^{d}h^{c}{}_{a}\right) .
\end{align*}
For the scalar field $\phi$, except the pure scalar interaction term $4 \lambda_s \phi^3 \bar{\phi}$, and the scalar and gravifield interactions are found to be,
\begin{align*}
  &\frac{1}{M_s \bar{a}(x)}\left(2 h^{ab} \partial_{a}\phi \partial_{b}h^{c}{}_{c} + 2 h^{ab} \partial_{a}h^{c}{}_{c} \partial_{b}\phi -  \phi \partial_{b}h^{c}{}_{c} \partial^{b}h^{a}{}_{a} -  \phi \partial_{b}h^{ab} \partial_{c}h_{a}{}^{c} \right.
  \\-& 2 h^{ab} \partial_{b}\phi \partial_{c}h_{a}{}^{c} - 2 h^{ab} \partial_{a}\phi \partial_{c}h_{b}{}^{c} + 2 \phi \partial^{b}h^{a}{}_{a} \partial_{c}h_{b}{}^{c} - 2 h^{ab} \partial_{b}h^{c}{}_{a} \partial_{c}\phi -  \phi \partial_{a}h_{bc} \partial^{c}h^{ab}
  \\-&\left.  \tfrac{1}{2} \phi \partial_{a}h_{cb} \partial^{c}h^{ab} + \tfrac{1}{2} \phi \partial_{b}h_{ac} \partial^{c}h^{ab} + \tfrac{1}{2} \phi \partial_{c}h_{ab} \partial^{c}h^{ab} + \tfrac{1}{2} \phi \partial_{c}h_{ba} \partial^{c}h^{ab} + 2 h^{ab} \partial_{c}\phi \partial^{c}h_{ba}\right) ,
\end{align*}
for $h-\phi-h$, and 
\begin{align*}
  \frac{\alpha_E}{M_W \bar{a}(x)}\left(4 \phi \partial_{a}\phi \partial_{b}h^{ab} + \frac{12\alpha_E-1}{\alpha_E} h^{ab} \partial_{a}\phi \partial_{b}\phi - 4 \phi \partial_{b}\phi \partial^{b}h^{a}{}_{a}\right)
\end{align*}
for $h-\phi-\phi$, as well as 
\begin{align*}
  \frac{1}{\sqrt{2\alpha_E}}(- g_w h^{ab} w_{b} \partial_{a}\phi -  g_w h^{ab} w_{a} \partial_{b}\phi)
\end{align*}
or $h-w-\phi$, and 
\begin{align*}
  &\frac{g_w\sqrt{\lambda/\alpha_E}M_s}{12\alpha_E\kappa_\mu\kappa^\mu} \bar{a}(x)\left(2 h^{ab} h_{b}{}^{c} \kappa_{c} w_{a} + 2 h^{ab} h_{b}{}^{c} \kappa_{a} w_{c} + 2 h_{a}{}^{c} h^{ab} \kappa_{b} w_{c} \right) ,
\end{align*}
for $h-h-w$. With coupling to the spin gauge field, we obtain 
\begin{align*}
  &\frac{g_h^2}{M_W\bar{a}(x)}\left(  g_1 h^{ab} \partial_{a}\Omega^{cdm} \partial_{b}\Omega_{cdm}-(g_1+g_2) h^{ab} \partial_{a}\Omega^{cdm} \partial_{b}\Omega_{dcm}-( g_1+g_2) h^{ab} \partial_{b}\Omega_{cdm} \partial^{m}\Omega_{a}{}^{cd}  \right.
  \\+&  g_4 h^{ab} \partial_{b}\Omega^{m}{}_{dm} \partial_{c}\Omega_{a}{}^{cd} + g_4 h^{ab} \partial_{a}\Omega^{m}{}_{dm} \partial_{c}\Omega_{b}{}^{cd} +  g_4 h^{ab} \partial_{b}\Omega^{c}{}_{a}{}^{d} \partial_{c}\Omega^{m}{}_{dm} + \tfrac{3}{4} g_4 h^{ab} \partial_{b}\Omega^{m}{}_{cm} \partial_{d}\Omega^{c}{}_{a}{}^{d}
  \\+&  \tfrac{1}{4} g_4 h^{ab} \partial_{a}\Omega^{m}{}_{cm} \partial_{d}\Omega^{c}{}_{b}{}^{d} - \tfrac{3}{4} g_4 h^{ab} \partial_{c}\Omega_{b}{}^{cd} \partial_{d}\Omega^{m}{}_{am} - \tfrac{1}{4} g_4 h^{ab} \partial_{c}\Omega_{a}{}^{cd} \partial_{d}\Omega^{m}{}_{bm}- g_4 h^{ab} \partial_{b}\Omega^{c}{}_{a}{}^{d} \partial_{d}\Omega^{m}{}_{cm}
  \\+&  g_4h^{ab} \partial_{c}\Omega^{m}{}_{dm} \partial^{d}\Omega_{ba}{}^{c} - g_4 h^{ab} \partial_{d}\Omega^{m}{}_{cm} \partial^{d}\Omega_{ba}{}^{c} 
- \tfrac{3}{4} g_4 h^{ab} \partial_{b}\Omega^{m}{}_{dm} \partial^{d}\Omega^{c}{}_{ac} - \tfrac{1}{4} g_4 h^{ab} \partial_{a}\Omega^{m}{}_{dm} \partial^{d}\Omega^{c}{}_{bc} 
\\-& g_4 h^{ab} \partial_{c}\Omega_{a}{}^{cd} \partial_{m}\Omega_{bd}{}^{m} + g_4 h^{ab} \partial_{b}\Omega^{c}{}_{a}{}^{d} \partial_{m}\Omega_{cd}{}^{m} - g_4 h^{ab} \partial^{d}\Omega_{ba}{}^{c} \partial_{m}\Omega_{cd}{}^{m}  +  \tfrac{3}{4} g_4 h^{ab} \partial_{c}\Omega_{b}{}^{cd} \partial_{m}\Omega_{da}{}^{m}
  \\+&  \tfrac{1}{4} g_4 h^{ab} \partial_{c}\Omega_{a}{}^{cd} \partial_{m}\Omega_{db}{}^{m} +  g_4 h^{ab} \partial_{b}\Omega^{c}{}_{a}{}^{d} \partial_{m}\Omega_{dc}{}^{m} +  g_4 h^{ab} \partial^{d}\Omega_{ba}{}^{c} \partial_{m}\Omega_{dc}{}^{m} 
  \\-& g_1 h^{ab} \partial_{b}\Omega_{mcd} \partial^{m}\Omega_{a}{}^{cd} -( g_1+g_2) h^{ab} \partial_{d}\Omega_{bcm} \partial^{m}\Omega_{a}{}^{cd} +  g_1 h^{ab} \partial_{m}\Omega_{bcd} \partial^{m}\Omega_{a}{}^{cd} 
 \\-&( g_1+g_2) h^{ab} \partial_{a}\Omega_{cdm} \partial^{m}\Omega_{b}{}^{cd}-g_1 h^{ab} \partial_{a}\Omega_{mcd} \partial^{m}\Omega_{b}{}^{cd} + (g_1+g_2) h^{ab} \partial_{b}\Omega_{cdm} \partial^{m}\Omega^{c}{}_{a}{}^{d}
  \\+& 2 g_2 h^{ab} \partial_{d}\Omega_{cam} \partial^{m}\Omega_{b}{}^{cd} -( g_1+g_2) h^{ab} \partial_{d}\Omega_{mac} \partial^{m}\Omega_{b}{}^{cd}-( g_1+g_2) h^{ab} \partial_{m}\Omega_{cad} \partial^{m}\Omega_{b}{}^{cd}  
  \\ -&\left. 2 g_2 h^{ab} \partial_{b}\Omega_{dcm} \partial^{m}\Omega^{c}{}_{a}{}^{d} +( g_1+g_2) h^{ab} \partial_{b}\Omega_{mcd} \partial^{m}\Omega^{c}{}_{a}{}^{d}- g_4 h^{ab} \partial_{a}\Omega^{c}{}_{c}{}^{d} \partial_{b}\Omega^{m}{}_{dm}  \right) ,
\end{align*}
for $h-\Omega-\Omega$.  More interactions include 
\begin{align*}
  \frac{1}{M_W \bar{a}(x)}\left(h^{ab} \partial_{a}w^{c} \partial_{b}w_{c} -  h^{ab} \partial_{b}w_{c} \partial^{c}w_{a} + h^{ab} \partial_{c}w_{b} \partial^{c}w_{a} -  h^{ab} \partial_{a}w_{c} \partial^{c}w_{b}\right)
\end{align*}
for $h-w-w$, and 
\begin{equation*}
  g_w w^{\nu} \phi \partial_{\nu}\phi ,
\end{equation*}
for $w-\phi-\phi$, as well as 
\begin{equation*}
  g_w^2 M_s \bar{a}(x) w^{\nu} w_{\nu}\phi ,
\end{equation*}
for $w-w-\phi$.
\section*{Appendix C: Tensor Projection Operators}

Here we show the exact expression of projection operators $(P_{f_1 f_2}^{J^P})$ for the spin gauge field, gravifield and scalar, in which we have used the definitions $\theta_{\mu\nu}=\eta_{\mu\nu}-\omega_{\mu\nu}$  and $\omega_{\mu\nu}=\tfrac{k_\mu k_\nu}{k^2}$ for short.
\begin{IEEEeqnarray*}{ll}
P_\Omega^{2^-}{}_{\mu ab,\nu cd}&=\frac{1}{6} [ 2 \theta^{\mu \nu} (\theta^{ac} \theta^{bd}  - \theta^{ad} \theta^{bc} ) + \theta^{a\nu} \theta^{bd} \theta^{\mu c} -  \theta^{ad} \theta^{b\nu} \theta^{\mu c} -  \theta^{a\nu} \theta^{bc} \theta^{\mu d} + \theta^{ac} \theta^{b\nu} \theta^{\mu d} ]\nonumber\\&-  \frac{1}{4}(\theta^{bd} \theta^{\mu a} \theta^{\nu c} -  \theta^{ad} \theta^{\mu b} \theta^{\nu c} -  \theta^{bc} \theta^{\mu a} \theta^{\nu d} + \theta^{ac} \theta^{\mu b} \theta^{\nu d})
\nonumber\\
P_\Omega^{2^+}{}_{\mu ab,\nu cd}&= \frac{1}{4} \theta^{\mu \nu} (\theta^{bd}  \omega^{ac} -  \theta^{bc}  \omega^{ad} -  \theta^{ad} \omega^{bc} + \theta^{ac} \theta^{\mu \nu} \omega^{bd}) \nonumber\\
& + \frac{1}{4} (\theta^{b\nu} \theta^{\mu d} \omega^{ac} -  \theta^{b\nu} \theta^{\mu c} \omega^{ad} -  \theta^{a\nu} \theta^{\mu d} \omega^{bc} + \theta^{a\nu} \theta^{\mu c} \omega^{bd}) \nonumber\\ &-  \frac{1}{6}\theta^{\mu b} \theta^{\nu d} \omega^{ac} -  \theta^{\mu b} \theta^{\nu c} \omega^{ad} -  \theta^{\mu a} \theta^{\nu d} \omega^{bc} + \theta^{\mu a} \theta^{\nu c} \omega^{bd}
\nonumber\\
P_h^{2^+}{}_{\mu a,\nu b}&=\frac{1}{2} \theta^{\mu \nu} \theta^{ab}  + \frac{1}{2} \theta^{a\nu} \theta^{\mu b}  -  \frac{1}{3}\theta^{\mu a} \theta^{\nu b}
\nonumber\\
P_{h\Omega}^{2^+}{}_{\mu a,\nu bc}&=\frac{p^{c}}{2 \sqrt{2p^2}} (\theta^{b\mu} \theta^{\nu a} -  \frac{2}{3} \theta^{\mu a} \theta^{\nu b} + \theta^{ba} \theta^{\nu \mu}) -  \frac{p^{b}}{2 \sqrt{2p^2}} (\theta^{c\mu} \theta^{\nu a} -  \frac{2}{3} \theta^{\mu a} \theta^{\nu c} + \theta^{ca} \theta^{\nu \mu})
\nonumber\\
P_{\Omega h}^{2^+}{}_{\mu ab,\nu c}&=\frac{p^{b}}{2 \sqrt{2p^2}} (\theta^{a\nu} \theta^{\mu c} + \theta^{ac} \theta^{\mu \nu} -  \frac{2}{3} \theta^{\mu a} \theta^{\nu c}) -  \frac{p^{a}}{2 \sqrt{2p^2}} (\theta^{b\nu} \theta^{\mu c} + \theta^{bc} \theta^{\mu \nu} -  \frac{2}{3} \theta^{\mu b} \theta^{\nu c})
\nonumber\\
P_{\Omega_1}^{1^+}{}_{\mu ab,\nu cd}&= \frac{1}{6} [ \theta^{\mu \nu}  (\theta^{bd} \omega^{ac}  -  \theta^{bc}  \omega^{ad} )  - \theta^{b\nu} \theta^{\mu d} \omega^{ac} + \theta^{b\nu} \theta^{\mu c} \omega^{ad} -  \theta^{bd} \theta^{\mu c} \omega^{a\nu} + \theta^{bc} \theta^{\mu d} \omega^{a\nu} \nonumber\\
&  +\theta^{\mu \nu} (\theta^{ac}  \omega^{bd} -  \theta^{ad} \omega^{bc}) + \theta^{a\nu} \theta^{\mu d} \omega^{bc}  -  \theta^{a\nu} \theta^{\mu c} \omega^{bd} + + \theta^{ad} \theta^{\mu c} \omega^{b\nu} -  \theta^{ac} \theta^{\mu d} \omega^{b\nu} \nonumber\\
&-  \theta^{a\nu} \theta^{bd} \omega^{\mu c} + \theta^{ad} \theta^{b\nu} \omega^{\mu c} + \theta^{a\nu} \theta^{bc} \omega^{\mu d} -  \theta^{ac} \theta^{b\nu} \omega^{\mu d} -  \theta^{ad} \theta^{bc} \omega^{\mu \nu} + \theta^{ac} \theta^{bd} \omega^{\mu \nu}]
\nonumber\\
P_{\Omega_2}^{1^+}{}_{\mu ab,\nu cd}&=\frac{1}{12} \bigl( \theta^{\mu \nu}  (\theta^{bd} \omega^{ac} -  \theta^{bc} \omega^{ad} ) - \theta^{b\nu} \theta^{\mu d} \omega^{ac}  + \theta^{b\nu} \theta^{\mu c} \omega^{ad}  + \theta^{a\nu} \theta^{\mu d} \omega^{bc} \nonumber\\
& + \theta^{\mu \nu} (\theta^{ac}  \omega^{bd} -  \theta^{ad} \omega^{bc}) -  \theta^{a\nu} \theta^{\mu c} \omega^{bd}  - 2 (- \theta^{bd} \theta^{\mu c} \omega^{a\nu} + \theta^{bc} \theta^{\mu d} \omega^{a\nu} \nonumber\\&+ \theta^{ad} \theta^{\mu c} \omega^{b\nu} -  \theta^{ac} \theta^{\mu d} \omega^{b\nu}) - 2 (- \theta^{a\nu} \theta^{bd} \omega^{\mu c} + \theta^{ad} \theta^{b\nu} \omega^{\mu c} + \theta^{a\nu} \theta^{bc} \omega^{\mu d} \nonumber\\&-  \theta^{ac} \theta^{b\nu} \omega^{\mu d}) + 4 (- \theta^{ad} \theta^{bc} \omega^{\mu \nu} + \theta^{ac} \theta^{bd} \omega^{\mu \nu})\bigr)
\nonumber\\
P_h^{1^+}{}_{\mu a,\nu b}&= \frac{1}{2} (  \theta^{\mu \nu} \theta^{ab}-  \theta^{a\nu} \theta^{b\mu} )
\nonumber\\
P_{\Omega_1h}^{1^+}{}_{\mu ab,\nu c}&=\frac{p^{\mu} \theta^{a\nu} \theta^{bc}}{2 \sqrt{3 p^2}} -  \frac{p^{\mu} \theta^{ac} \theta^{b\nu}}{2 \sqrt{3 p^2}} -  \frac{p^{b} \theta^{a\nu} \theta^{c\mu}}{2 \sqrt{3 p^2}} + \frac{p^{a} \theta^{b\nu} \theta^{c\mu}}{2 \sqrt{3 p^2}} + \frac{p^{b} \theta^{ac} \theta^{\mu \nu}}{2 \sqrt{3 p^2}} -  \frac{p^{a} \theta^{bc} \theta^{\mu \nu}}{2 \sqrt{3 p^2}}
\nonumber\\
P_{h\Omega_1}^{1^+}{}_{\mu a,\nu bc}&=\frac{p^{\nu} \theta^{ac} \theta^{b\mu}}{2 \sqrt{3 p^2}} -  \frac{p^{c} \theta^{a\nu} \theta^{b\mu}}{2 \sqrt{3 p^2}} -  \frac{p^{\nu} \theta^{ab} \theta^{c\mu}}{2 \sqrt{3 p^2}} + \frac{p^{b} \theta^{a\nu} \theta^{c\mu}}{2 \sqrt{3 p^2}} + \frac{p^{c} \theta^{ab} \theta^{\mu \nu}}{2 \sqrt{3 p^2}} -  \frac{p^{b} \theta^{ac} \theta^{\mu \nu}}{2 \sqrt{3 p^2}}
\nonumber\\
P_{\Omega_2h}^{1^+}{}_{\mu ab,\nu c}&=\frac{p^{\mu} \theta^{a\nu} \theta^{bc}}{\sqrt{6 p^2}} -  \frac{p^{\mu} \theta^{ac} \theta^{b\nu}}{\sqrt{6 p^2}} + \frac{p^{b} \theta^{a\nu} \theta^{c\mu}}{2 \sqrt{6 p^2}} -  \frac{p^{a} \theta^{b\nu} \theta^{c\mu}}{2 \sqrt{6 p^2}} -  \frac{p^{b} \theta^{ac} \theta^{\mu \nu}}{2 \sqrt{6 p^2}} + \frac{p^{a} \theta^{bc} \theta^{\mu \nu}}{2 \sqrt{6 p^2}}
\nonumber\\
P_{h\Omega_2}^{1^+}{}_{\mu a,\nu bc}&=\frac{p^{\nu} \theta^{ac} \theta^{b\mu}}{\sqrt{6 p^2}} + \frac{p^{c} \theta^{a\nu} \theta^{b\mu}}{2 \sqrt{6 p^2}} -  \frac{p^{\nu} \theta^{ab} \theta^{c\mu}}{\sqrt{6 p^2}} -  \frac{p^{b} \theta^{a\nu} \theta^{c\mu}}{2 \sqrt{6 p^2}} -  \frac{p^{c} \theta^{ab} \theta^{\mu \nu}}{2 \sqrt{6 p^2}} + \frac{p^{b} \theta^{ac} \theta^{\mu \nu}}{2 \sqrt{6 p^2}}
\nonumber\\
P_{\Omega_1\Omega_2}^{1^+}{}_{\mu ab,\nu cd}&=\frac{1}{6 \sqrt{2}}(\theta^{b\nu} \theta^{\mu d} \omega^{ac} -  \theta^{bd} \theta^{\mu \nu} \omega^{ac} -  \theta^{b\nu} \theta^{\mu c} \omega^{ad} + \theta^{bc} \theta^{\mu \nu} \omega^{ad} - 2 \theta^{bd} \theta^{\mu c} \omega^{a\nu} \nonumber\\&+ 2 \theta^{bc} \theta^{\mu d} \omega^{a\nu} -  \theta^{a\nu} \theta^{\mu d} \omega^{bc} + \theta^{ad} \theta^{\mu \nu} \omega^{bc} + \theta^{a\nu} \theta^{\mu c} \omega^{bd} -  \theta^{ac} \theta^{\mu \nu} \omega^{bd} \nonumber\\&+ 2 \theta^{ad} \theta^{\mu c} \omega^{b\nu} - 2 \theta^{ac} \theta^{\mu d} \omega^{b\nu} + \theta^{a\nu} \theta^{bd} \omega^{\mu c} -  \theta^{ad} \theta^{b\nu} \omega^{\mu c} -  \theta^{a\nu} \theta^{bc} \omega^{\mu d} \nonumber\\&+ \theta^{ac} \theta^{b\nu} \omega^{\mu d} - 2 \theta^{ad} \theta^{bc} \omega^{\mu \nu} + 2 \theta^{ac} \theta^{bd} \omega^{\mu \nu})
\nonumber\\
P_{\Omega_2\Omega_1}^{1^+}{}_{\mu ab,\nu cd}&=\frac{1}{6 \sqrt{2}}\bigl(\theta^{b\nu} \theta^{\mu d} \omega^{ac} -  \theta^{bd} \theta^{\mu \nu} \omega^{ac} -  \theta^{b\nu} \theta^{\mu c} \omega^{ad} + \theta^{bc} \theta^{\mu \nu} \omega^{ad} + \theta^{bd} \theta^{\mu c} \omega^{a\nu} \nonumber\\&-  \theta^{bc} \theta^{\mu d} \omega^{a\nu} -  \theta^{a\nu} \theta^{\mu d} \omega^{bc} + \theta^{ad} \theta^{\mu \nu} \omega^{bc} + \theta^{a\nu} \theta^{\mu c} \omega^{bd} -  \theta^{ac} \theta^{\mu \nu} \omega^{bd} \nonumber\\&-  \theta^{ad} \theta^{\mu c} \omega^{b\nu} + \theta^{ac} \theta^{\mu d} \omega^{b\nu} - 2 (\theta^{a\nu} \theta^{bd} \omega^{\mu c} -  \theta^{ad} \theta^{b\nu} \omega^{\mu c} -  \theta^{a\nu} \theta^{bc} \omega^{\mu d} \nonumber\\&+ \theta^{ac} \theta^{b\nu} \omega^{\mu d}) - 2 \theta^{ad} \theta^{bc} \omega^{\mu \nu} + 2 \theta^{ac} \theta^{bd} \omega^{\mu \nu}\bigr)
\nonumber\\
P_{\Omega_1}^{1^-}{}_{\mu ab,\nu cd}&=\frac{1}{4}(\theta^{bd} \theta^{\mu a} \theta^{\nu c} -  \theta^{ad} \theta^{\mu b} \theta^{\nu c} -  \theta^{bc} \theta^{\mu a} \theta^{\nu d} + \theta^{ac} \theta^{\mu b} \theta^{\nu d})
\nonumber\\
P_{\Omega_2}^{1^-}{}_{\mu ab,\nu cd}&=\frac{1}{2} (\theta^{bd} \omega^{ac} \omega^{\mu \nu} -  \theta^{bc} \omega^{ad} \omega^{\mu \nu} -  \theta^{ad} \omega^{bc} \omega^{\mu \nu} + \theta^{ac} \omega^{bd} \omega^{\mu \nu})
\nonumber\\
P_{h_1}^{1^-}{}_{\mu a,\nu b}&=\frac{1}{2} (\theta^{\mu \nu} \omega^{ab} -  \theta^{\mu b} \omega^{a\nu} -  \theta^{a\nu} \omega^{\mu b} + \theta^{ab} \omega^{\mu \nu})
\nonumber\\
P_{h_2}^{1^-}{}_{\mu a,\nu b}&=\frac{1}{2} (\theta^{\mu \nu} \omega^{ab} + \theta^{\mu b} \omega^{a\nu} + \theta^{a\nu} \omega^{\mu b} + \theta^{ab} \omega^{\mu \nu})
\nonumber\\
P_{h_1h_2}^{1^-}{}_{\mu a,\nu b}&=\frac{1}{2} (\theta^{\mu \nu} \omega^{ab} + \theta^{\mu b} \omega^{a\nu} -  \theta^{a\nu} \omega^{\mu b} -  \theta^{ab} \omega^{\mu \nu})
\nonumber\\
P_{h_2h_1}^{1^-}{}_{\mu a,\nu b}&=\frac{1}{2} (\theta^{\mu \nu} \omega^{ab} -  \theta^{\mu b} \omega^{a\nu} + \theta^{a\nu} \omega^{\mu b} -  \theta^{ab} \omega^{\mu \nu})
\nonumber\\
P_{h_1\Omega_1}^{1^-}{}_{\mu a,\nu bc}&=-\frac{1}{2 \sqrt{2}}(\frac{p^{\mu}}{\sqrt{p^2}} \theta^{ca} \theta^{\nu b} -  \frac{p^{a}}{\sqrt{p^2}} \theta^{c\mu} \theta^{\nu b} -  \frac{p^{\mu}}{\sqrt{p^2}} \theta^{ba} \theta^{\nu c} + \frac{p^{a}}{\sqrt{p^2}} \theta^{b\mu} \theta^{\nu c})
\nonumber\\
P_{h_1\Omega_2}^{1^-}{}_{\mu a,\nu bc}&=-\frac{1}{2} (\frac{p^{\mu}}{\sqrt{p^2}}  \theta^{ca} \omega^{\nu b}-  \frac{p^{a}}{\sqrt{p^2}} \theta^{c\mu} \omega^{\nu b} -  \frac{p^{\mu}}{\sqrt{p^2}} \theta^{ba} \omega^{\nu c} + \frac{p^{a}}{\sqrt{p^2}} \theta^{b\mu} \omega^{\nu c})
\nonumber\\
P_{\Omega_1h_1}^{1^-}{}_{\mu ab,\nu c}&=-\frac{1}{2 \sqrt{2}}(\frac{p^{\nu}}{\sqrt{p^2}} \theta^{bc} \theta^{\mu a} -  \frac{p^{c}}{\sqrt{p^2}} \theta^{b \nu} \theta^{\mu a} -  \frac{p^{\nu}}{\sqrt{p^2}} \theta^{ac} \theta^{\mu b} + \frac{p^{c}}{\sqrt{p^2}} \theta^{a \nu} \theta^{\mu b})
\nonumber\\
P_{\Omega_2h_1}^{1^-}{}_{\mu ab,\nu c}&=-\frac{p^{\nu}}{2\sqrt{p^2}} \theta^{bc} \omega^{\mu a} +  \frac{p^{c}}{2\sqrt{p^2}} \theta^{b\nu} \omega^{\mu a} +  \frac{p^{\nu}}{2\sqrt{p^2}} \theta^{ac} \omega^{\mu b} - \frac{p^{c}}{2\sqrt{p^2}} \theta^{a\nu} \omega^{\mu b}
\nonumber\\
P_{h_2\Omega_1}^{1^-}{}_{\mu a,\nu bc}&=\frac{1}{2 \sqrt{2}}(\frac{p^{\mu}}{\sqrt{p^2}} \theta^{ca} \theta^{\nu b} +  \frac{p^{a}}{\sqrt{p^2}} \theta^{c\mu} \theta^{\nu b} -  \frac{p^{\mu}}{\sqrt{p^2}} \theta^{ba} \theta^{\nu c} - \frac{p^{a}}{\sqrt{p^2}} \theta^{b\mu} \theta^{\nu c})
\nonumber\\
P_{h_2\Omega_2}^{1^-}{}_{\mu a,\nu bc}&=\frac{1}{2} (\frac{p^{\mu}}{\sqrt{p^2}} \theta^{ca} \omega^{\nu b} +  \frac{p^{a}}{\sqrt{p^2}} \theta^{c\mu} \omega^{\nu b} -  \frac{p^{\mu}}{\sqrt{p^2}} \theta^{ba} \omega^{\nu c} - \frac{p^{a}}{\sqrt{p^2}} \theta^{b\mu} \omega^{\nu c})
\nonumber\\
P_{\Omega_1h_2}^{1^-}{}_{\mu ab,\nu c}&=\frac{1}{2 \sqrt{2}}(\frac{p^{\nu}}{\sqrt{p^2}} \theta^{bc} \theta^{\mu a} +  \frac{p^{c}}{\sqrt{p^2}} \theta^{b \nu} \theta^{\mu a} -  \frac{p^{\nu}}{\sqrt{p^2}} \theta^{ac} \theta^{\mu b} - \frac{p^{c}}{\sqrt{p^2}} \theta^{a \nu} \theta^{\mu b})
\nonumber\\
P_{\Omega_2h_2}^{1^-}{}_{\mu ab,\nu c}&=\frac{p^{\nu}}{2\sqrt{p^2}} \theta^{bc} \omega^{\mu a} +  \frac{p^{c}}{2\sqrt{p^2}} \theta^{b\nu} \omega^{\mu a} -  \frac{p^{\nu}}{2\sqrt{p^2}} \theta^{ac} \omega^{\mu b} - \frac{p^{c}}{2\sqrt{p^2}} \theta^{a\nu} \omega^{\mu b}
\nonumber\\
P_{\Omega_1\Omega_2}^{1^-}{}_{\mu ab,\nu cd}&=\frac{1}{2 \sqrt{2}}(\theta^{bd} \theta^{\mu a} \omega^{\nu c} -  \theta^{ad} \theta^{\mu b} \omega^{\nu c} -  \theta^{bc} \theta^{\mu a} \omega^{\nu d} + \theta^{ac} \theta^{\mu b} \omega^{\nu d})
\nonumber\\
P_{\Omega_2\Omega_1}^{1^-}{}_{\mu ab,\nu cd}&=\frac{1}{2 \sqrt{2}}(\theta^{bd} \theta^{\nu c} \omega^{\mu a} -  \theta^{bc} \theta^{\nu d} \omega^{\mu a} -  \theta^{ad} \theta^{\nu c} \omega^{\mu b} + \theta^{ac} \theta^{\nu d} \omega^{\mu b})
\nonumber\\
P_\Omega^{0^-}{}_{\mu ab,\nu cd}&= \frac{1}{6} (- \theta^{a\nu} \theta^{bd} \theta^{\mu c} + \theta^{ad} \theta^{b\nu} \theta^{\mu c} + \theta^{a\nu} \theta^{bc} \theta^{\mu d} -  \theta^{ac} \theta^{b\nu} \theta^{\mu d}\nonumber\\& -  \theta^{ad} \theta^{bc} \theta^{\mu \nu} + \theta^{ac} \theta^{bd} \theta^{\mu \nu})
\nonumber\\
P_\Omega^{0^+}{}_{\mu ab,\nu cd}&=\frac{1}{6}(\theta^{\mu b} \theta^{\nu d} \omega^{ac} -  \theta^{\mu b} \theta^{\nu c} \omega^{ad} -  \theta^{\mu a} \theta^{\nu d} \omega^{bc} + \theta^{\mu a} \theta^{\nu c} \omega^{bd})
\nonumber\\
P_{h_1}^{0^+}{}_{\mu a,\nu b}&=\frac{1}{3} \theta^{a\mu} \theta^{b\nu}
\nonumber\\
P_{h_2}^{0^+}{}_{\mu a,\nu b}&=\omega^{a\mu} \omega^{b\nu}
\nonumber\\
P_\phi^{0^+}&=1
\nonumber\\
P_{h_1h_2}^{0^+}{}_{\mu a,\nu b}&=\frac{1}{\sqrt{3}}\theta^{a\mu} \omega^{b\nu}
\nonumber\\
P_{h_2h_1}^{0^+}{}_{\mu a,\nu b}&=\frac{1}{\sqrt{3}}\theta^{b\nu} \omega^{a\mu}
\nonumber\\
P_{h_1\phi}^{0^+}{}_{\mu a,}&=P_{\phi h_1}^{0^+}{}_{,\mu a}=\frac{\theta^{a\mu}}{\sqrt{3}}
\nonumber\\
P_{h_2\phi}^{0^+}{}_{\mu a,}&=P_{\phi h_2}^{0^+}{}_{,\mu a}=\omega^{a\mu}
\nonumber\\
P_{\Omega h_1}^{0^+}{}_{\mu ab,\nu c}&=- \frac{p^{b}}{3\sqrt{2p^2} } \theta^{\mu a} \theta^{\nu c} + \frac{p^{a}}{3\sqrt{2p^2} } \theta^{\mu b} \theta^{\nu c}
\nonumber\\
P_{\Omega h_2}^{0^+}{}_{\mu ab,\nu c}&=- \frac{p^{b}}{\sqrt{6p^2}} \theta^{\mu a} \omega^{\nu c} + \frac{p^{a}}{\sqrt{6p^2}} \theta^{\mu b} \omega^{\nu c}
\nonumber\\
P_{h_1\Omega }^{0^+}{}_{\mu a,\nu bc}&=- \frac{p^{c}}{3\sqrt{2p^2} } \theta^{\nu b} \theta^{\mu a} + \frac{p^{b}}{3\sqrt{2p^2} } \theta^{\mu a} \theta^{\nu c}
\nonumber\\
P_{h_2\Omega }^{0^+}{}_{\mu a,\nu bc}&=- \frac{p^{c}}{\sqrt{6p^2}} \theta^{\nu b} \omega^{\mu a} + \frac{p^{b}}{\sqrt{6p^2}} \theta^{\nu c} \omega^{\mu a}
\nonumber\\
P_{\Omega \phi}^{0^+}{}_{\mu ab,}&=P_{\phi\Omega}^{0^+}{}_{,\mu ab}=- \frac{p^{b}}{\sqrt{6 p^2}} \theta^{\mu a} + \frac{p^{a}}{\sqrt{6 p^2}} \theta^{\mu b}
\end{IEEEeqnarray*}


\begin{thebibliography}{99}
\vspace*{-1mm}
\begin{small}\baselineskip=10pt\itemsep-2pt
\bibitem{Wu:2015wwa}
Y. L. Wu, \textit{Quantum field theory of gravity with spin and scaling gauge
  invariance and spacetime dynamics with quantum inflation}, Phys. Rev. D\textbf{93} (2016) 024012.

\bibitem{Wu:2015hoa}
Y. L. Wu, \textit{Theory of Quantum Gravity Beyond Einstein and Space-time
  Dynamics with Quantum Inflation},
  Int. J. Mod. Phys. A
  \textbf{30} (2015) 1545002.

\bibitem{Einstein:1915by}
A.~Einstein, \textit{On the General Theory of Relativity}, Sitzungsber.
  Preuss. Akad. Wiss. Berlin (Math. Phys.) \textbf{1915} (1915) 778.

\bibitem{Voronov:1973kga}
N.~A. Voronov, \textit{Gravitational Compton effect and photoproduction of
  gravitons by electrons}, Sov. Phys. JETP {\bfseries 37} (1973) 953.

\bibitem{Choi:1994ax}
S.~Y. Choi, J.~S. Shim and H.~S. Song, \textit{Factorization and polarization in
  linearized gravity}, Phys. Rev. D \textbf{
  51} (1995) 2751.

\bibitem{Kerlick:1975tr}
G.~D. Kerlick, \textit{Cosmology and Particle Pair Production via Gravitational
  Spin Spin Interaction in the Einstein-Cartan-Sciama-Kibble Theory of
  Gravity}, Phys. Rev. D
  \textbf{12} (1975) 3004.

\bibitem{Hehl:1976kj}
F.~W. Hehl, P.~Von Der~Heyde, G.~D. Kerlick and J.~M. Nester, \textit{General
  Relativity with Spin and Torsion: Foundations and Prospects},
  Rev. Mod. Phys.
  \textbf{48} (1976) 393.

\bibitem{Hehl:1974cn}
F.~W. Hehl, G.~D. Kerlick and P.~Von Der~Heyde, \textit{General relativity with
  spin and torsion and its deviations from Einstein's theory},   {Phys. Rev. D \textbf{
  10} (1974) 1066.

\bibitem{VanNieuwenhuizen:1973fi}
P.~Van~Nieuwenhuizen, \textit{On ghost-free tensor lagrangians and linearized
  gravitation}, Nucl. Phys. B
  \textbf{60} (1973) 478.

\bibitem{Rivers:1964rj}
R.~Rivers, , {Nuovo Cim.} \textbf{34} (1964) 387.

\bibitem{Sezgin:1979zf}
E.~Sezgin and P.~van Nieuwenhuizen, \textit{New Ghost Free Gravity Lagrangians
  with Propagating Torsion},
   Phys. Rev. D \textbf{
  21} (1980) 3269.

\bibitem{Sezgin:1980tp}
E.~Sezgin and P.~van Nieuwenhuizen, \textit{{Renormalizability Properties of
  Antisymmetric Tensor Fields Coupled to Gravity}},
   Phys. Rev. D \textbf{
  22} (1980) 301.

\bibitem{Wu:2017urh}
Y. L. Wu, \textit{{Hyperunified field theory and gravitational gauge-geometry
  duality}}, {Eur.
  Phys. J.} \textbf{C78} (2018) 28}
  .

\bibitem{Ganjali:2018hyj}
M.~A. Ganjali, V.~Amirkhani and A.~ShamlouMehr, \textit{{Non-Relativistic
  Fermion-Fermion Scattering in Higher Derivative Gravity}}, {\ttfamily hep-th/1802.03470}.

\bibitem{Peskin:1995ev}
M.~E. Peskin and D.~V. Schroeder, \textit{{An Introduction to quantum field
  theory}}. Addison-Wesley, Reading, USA, 1995.

\bibitem{Schwartz:2013pla}
M.~D. Schwartz, \textit{{Quantum Field Theory and the Standard Model}}. Cambridge
  University Press, 2014.
\end{small}
\end{thebibliography}

\end{document}